\documentclass[sigconf, screen, nonacm]{acmart}

\usepackage{amsmath}
\usepackage{algorithmic}
\usepackage{graphicx}
\usepackage{textcomp}
\usepackage{xcolor}

\usepackage{tikz}
\usepackage{pgfplots}
\pgfplotsset{compat=1.18}
\usepgfplotslibrary{groupplots}

\usepackage{filecontents}
\usepackage{xspace}
\usepackage{balance}
\usepackage{enumitem}
\usepackage[ruled,lined,linesnumbered,noend]{algorithm2e}
\usepackage{cleveref}
\usepackage{multirow}
\usepackage{afterpage}
\usepackage{siunitx}
\usepackage{tabularx}
\usepackage{booktabs}
\newcolumntype{Y}{>{\raggedleft\arraybackslash}X}
\usepackage{listings}
\usepackage{subcaption}
\usepackage{threeparttable}
\usepackage{caption}
\usepackage{placeins}
\usepackage{afterpage}
\usepackage{amsthm}

\usepackage{float}
\usepackage{bm}

\usepackage[table]{xcolor}
\usepackage{arydshln}

\usepackage{listings}

\newcommand{\ourname}{\textsc{Gutenberg}\xspace}

\definecolor{forestgreen}{rgb}{0.13, 0.55, 0.13}

\renewcommand{\emph}{\textit}

\newcommand{\RNum}[1]{\expandafter{\romannumeral #1\relax}}
\newcommand\romenum[1]{\mbox{(\textit{\RNum{#1}})}\nolinebreak{}}

\newcommand{\refsep}{\,}
\newcommand{\secref}[1]{\S\ref{#1}}
\newcommand{\figref}[1]{Fig.\refsep\ref{#1}}
\newcommand{\tabref}[1]{Table\refsep\ref{#1}}

\newcommand{\myparagraph}[1]{\smallskip\noindent {\bf #1.}}

\newcommand{\cpundp}{CPU+NDP\xspace}

\SetKwProg{Struct}{Struct}{}{end}

\begin{document}

\title{\ourname: Taming Latency-Critical Online Datacenter Services with Near-Data-Processing}

\author{Qiushi Lin}
\affiliation{\institution{University of Texas at Austin}\city{Austin}\state{Texas}\country{USA}}
\author{Phillip B. Gibbons}
\affiliation{\institution{Carnegie Mellon University}\city{Pittsburgh}\state{Pennsylvania}\country{USA}}
\author{Jovan Stojkovic}
\affiliation{\institution{University of Texas at Austin}\city{Austin}\state{Texas}\country{USA}}
\author{Yiwei Zhao}
\affiliation{\institution{Indiana University Bloomington}\city{Bloomington}\state{Indiana}\country{USA}}

\begin{abstract}

Latency-critical online datacenter services place growing pressure on memory while requiring isolation, fairness, and predictable QoS.
Servers that pair a host CPU with near-data processing (NDP) units can relieve this pressure by executing requests next to memory, but three problems stand in the way.
First, prior NDP schedulers often optimize solely for throughput, ignoring tail-latency SLOs.
Second, uneven and shifting compute and memory traffic can overload a few NDP units and increase tail latency.
Third, when the CPU and an NDP unit share writable hot data, keeping these data coherent on every write significantly increases service times on both sides.

We present \ourname, a management system that combines the complementary strengths of the host CPU and the NDP units.
A lightweight sketch tracks how often each group of co-accessed data is requested and written, and \ourname uses these statistics at two levels.
At the request level, SLO-aware routing sends each request to the CPU or to an NDP unit according to its remaining deadline.
At the memory level, \ourname manages data at the granularity of hot groups: it balances work across the NDP units and the CPU, keeps write-heavy data on a single side, and guards against shifts in the read/write mix, the data placement, and the bottleneck.
Per-tenant pacing isolates co-located tenants.

Cycle-accurate simulation results show that \ourname improves \textit{service capacity}---throughput while meeting tail-latency SLOs---by up to 65\% (28\% on geometric mean) over the best baselines across five services and three NDP hardware regimes.
Under a shifting workload it misses the SLO for only 0.04\% of requests, compared to at least 1.85\% for baselines, and it sustains two co-located services at 50\% of their standalone load, versus 30\% for the baselines.

\end{abstract}

\maketitle

\section{Introduction}
\label{sec:intro}

Many online datacenter services are highly latency sensitive, including
in-memory key-value stores and caches~\cite{mao2012masstree,dragonflydb,nishtala2013scaling,yang2021large,decandia2007dynamo,chen2025rearchitecting},
transaction processing over relational and financial-graph data~\cite{tu2013silo,kemper2011hyper,qi2025finbench,lim2017cicada,qiu2018real,huang2025tigon},
search and ranking~\cite{kasture2016tailbench,brin1998pagerank,hsu2015adrenaline,haque2015few,mohoney2025quake},
and machine-learning inference services spanning translation, speech, and image and media processing~\cite{kasture2016tailbench,choi2025hera,gupta2020recommendation,moina2024cloudvideo,jouppi2017datacenter,xia2025modm}.
Such \textit{latency-critical} services store data in memory (accessing SSDs would be too slow) and are subject to strict service-level objectives (SLOs) on the tail latency of request processing.
A key metric in systems supporting these services is \textbf{\textit{service capacity}}, defined as the highest load a server can sustain while meeting the service's tail-latency SLO, often reported as \textit{throughput under an SLO}~\cite{kaffes2019shinjuku,li2020pegasus}.
Increasing service capacity places significant pressure on the server's memory system.
Consequently, request latency and throughput are often bottlenecked by the latency and limited bandwidth of memory access, known as the \textit{memory wall} problem.

\providecommand{\cmark}{\textcolor{green!50!black}{$\checkmark$}}
\providecommand{\xmark}{\textcolor{red!75!black}{$\times$}}
\begin{table}[t]
  \centering
  \caption{\ourname vs.~Prior Systems.}
  \label{tab:intro-compare}
  \vspace{-0.15in}
  \scriptsize
  \setlength{\tabcolsep}{2pt}
  \begin{tabularx}{\columnwidth}{@{}>{\raggedright\arraybackslash}p{.16\columnwidth}>{\raggedright\arraybackslash}p{.1\columnwidth}*{4}{>{\centering\arraybackslash}X}@{}}
    \toprule
    System & Platform & SLO-Aware & Hot-Spot Relief & Coherence-Aware & CPU-Assisted NDP Execution \\
    \midrule
    Heracles~\cite{lo2015heracles} & Multicore & \cmark & \xmark & \xmark & N/A \\
    PARTIES~\cite{chen2019parties} & Multicore & \cmark & \xmark & \xmark & N/A \\
    Shinjuku~\cite{kaffes2019shinjuku} & Multicore & \cmark & \cmark & \xmark & N/A \\
    NetCache~\cite{jin2017netcache} & Dist.~Sys. & \xmark & \cmark & \xmark & N/A \\
    Pegasus~\cite{li2020pegasus} & Dist.~Sys. & \cmark & \cmark & \cmark & N/A \\
    PEI~\cite{ahn2015piminstructions} & NDP & \xmark & \xmark & \cmark & \cmark \\
    PIMCloud~\cite{chen2022pimcloud} & NDP & \cmark & \xmark & \xmark & \cmark \\
    ABNDP~\cite{tian2023abndp} & NDP & \xmark & \cmark & \xmark & \xmark \\
    \midrule
    \textbf{\ourname} (Ours) & NDP & \cmark & \cmark & \cmark & \cmark \\
    \bottomrule
  \end{tabularx}
  \vspace{-0.25in}
\end{table}

Near-data processing (NDP), or processing-in-memory (PIM), has recently reemerged as a promising architectural solution to the memory wall problem.
By integrating compute units (\textit{NDP cores}) near or within memory modules, NDP enables computation to be performed close to where data reside.
Instead of incurring the high cost of fetching the needed data from the module back to the CPU for processing (as in traditional von Neumann architectures), the processing occurs at the NDP core near or within the module holding the data.  
This design improves both performance and energy efficiency by leveraging the NDP core's low-latency, high-bandwidth, low-energy access to the module.
Enabled by advances in fabrication such as 3D stacking~\cite{jeddeloh2012hybriddram,lee2014hbm}, the past decade has seen a growing body of NDP research (e.g., see the references of~\cite{mutlu2023primer}) as well as the emergence of real-world industrial NDP products and prototypes~\cite{upmem,kwon2021samsunghbmpim,skhynix2022gddr6aim,xcena2026mx1}.
Most NDP work, however, targets batch workloads~\cite{ahn2015tesseract,gomez2022benchmarking} and is ill-suited to the streaming requests of online datacenter services.

To exploit efficient local memory access in NDP, computation should be offloaded to the \emph{NDP stack} (a memory stack with its own NDP cores) that holds its data.
However, data accesses in production workloads often skew towards a few hot items, whose popularity shifts over time~\cite{atikoglu2012workload,cao2020rocksdb,yang2020twitter}.
Under a direct offloading policy, NDP stacks holding hot data saturate quickly while other NDP stacks and the host CPU remain underutilized.
A central challenge for \cpundp systems is therefore to balance load across all compute units while limiting remote accesses and coherence traffic, a goal prior NDP work pursues through data partitioning, offloading decisions and coherence support~\cite{hsieh2016tom,zhang2018graphp,kang2022pimtreevldb,tian2023abndp,boroumand2019conda}.

For online services, meeting this goal raises four challenges that must be addressed (\tabref{tab:intro-compare}).
The table illustrates how prior work for NDP, as well as for more traditional multicore CPU and distributed systems settings, addresses only some of these challenges.

\romenum{1} \textbf{SLO-Aware:} Optimizing solely for throughput can result in high SLO violations and suboptimal service capacity.
Most NDP systems, including PEI and ABNDP, target batch workloads and optimize throughput~\cite{ahn2015tesseract,ahn2015piminstructions,tian2023abndp,gomez2022benchmarking}. In the distributed systems setting, NetCache decides which keys to cache solely by popularity~\cite{jin2017netcache}.

\romenum{2} \textbf{Hot-Spot Relief:} The read/write mix and the hot set of an online service keep shifting, so load must move off a hot stack as requests arrive, within a latency budget too tight to amortize data movement over a batch.
PEI chooses where each operation runs by data locality, ignoring skew.
NDP approaches that handle skew, such as PIM-tree~\cite{kang2022pimtreevldb} and ABNDP~\cite{tian2023abndp}, schedule each batch with its contention known in advance. 
Heracles and PARTIES allocate resources per service~\cite{lo2015heracles,chen2019parties}, and PIMCloud per application~\cite{chen2022pimcloud}, so none of them relieves a hot spot within one service.

\romenum{3} \textbf{Coherence-Aware:} Relieving a hot stack means serving its data elsewhere, which turns writable data into \textit{shared} data.
The CPU and an NDP stack share data by cache line but have no cache in common, so every synchronization between them goes through DRAM and the interconnect.
Our characterization (\secref{subsection:bg-observations}) shows that when both sides serve the same write-heavy group, their service times greatly increase.
Multicore schedulers such as Heracles, PARTIES and Shinjuku treat cores as interchangeable and do not account for the cost of sharing writable data among them, which a shared last-level cache keeps small.
Distributed servers share no cache lines: Pegasus keeps replicas of hot objects coherent with a directory in the switch~\cite{li2020pegasus}, and NetCache invalidates its cached keys on writes~\cite{jin2017netcache}.
Among NDP systems, LazyPIM and CoNDA make coherence cheaper for offloaded kernels~\cite{boroumand2016lazypim,boroumand2019conda}, ABNDP avoids write sharing by caching only data that stay read-only within a batch step~\cite{tian2023abndp}, and PIMCloud replicates only read-only pages~\cite{chen2022pimcloud}; none of them weighs this cost when deciding where a request runs.

\romenum{4} \textbf{CPU-Assisted NDP Execution:} Prior NDP systems (e.g., ABNDP) often use the host CPU mainly for dispatch, leaving its execution capacity underused~\cite{ahn2015tesseract,tian2023abndp}.
Using it to relieve hot stacks requires accounting for its stronger cores and larger caches, remote access to NDP data, the core budget it shares between dispatch and execution, and, for write-heavy data, the coherence cost above.
Load-balancing policies that treat all cores as interchangeable~\cite{kaffes2019shinjuku} or model heterogeneity through fixed relative speeds do not capture these workload-dependent tradeoffs.

\myparagraph{Our Solution: \ourname}
We present \ourname, a management system for \cpundp servers that addresses all four challenges, via techniques that combine the complementary strengths of the host CPU and the NDP stacks.
A lightweight \textit{Footprint Sketch} samples requests to track how often each group of co-accessed data is requested and written, and \ourname uses these statistics at two levels.
At the request level, Short-Term Steering (STS) routes each request with its remaining deadline in mind:
the request stays on its group's NDP stack as long as that stack can finish it in time, and otherwise runs on a spare core of the CPU.
STS also caps how many requests of a group run at once, below the point where contention on the group's locks takes over, and stops sending a write-heavy group's requests away from its stack once sharing the group starts to slow it down.
At the memory level, Long-Term Placement (LTP) manages hot data in groups.
Once per control period, it reassigns hot groups on overloaded stacks to a cold stack or to the CPU, keeping each write-heavy group on one side at a time (minimizing coherence).
Its \textit{Drift Guards} respond to specific shifts as the bottleneck moves: they keep write-heavy groups stable while the read/write mix changes and migrate a group to the stack that now serves it.
They also rebalance the number of host cores executing requests vs.~simply dispatching requests as the bottleneck moves.
Per-tenant pacing keeps one tenant's burst within its load share.

We evaluate \ourname in ZSim on five services, built on Masstree, Silo TPC-C and Redis, and on four TailBench services, across three NDP hardware regimes.
\ourname increases service capacity over the best baseline by up to 65\%, 28\% on geometric mean.
When the read/write mix, the hot set and the available CPU change during a run, it misses the SLO latency target for only 0.04\% of requests, compared to at least 1.85\% for the baselines.
\ourname serves two co-located services at 50\% of their standalone load, versus only 30\% for PIMCloud, and per-tenant pacing cuts the other tenants' SLO misses during a burst from 46\% to 15\%.

In summary, the contributions are:
\begin{itemize}[leftmargin=*]
  \item a characterization of demand and capacity on \cpundp servers under an SLO, including the cost of sharing write-heavy data between the CPU and the NDP stacks (\secref{subsection:bg-observations});
  \item \ourname, a management system that routes each request with its deadline in mind (STS) and manages hot data in groups, with drift guards for workload shifts (LTP) (\secref{subsection:design-fast}, \secref{subsection:design-slow});
  \item the Footprint Sketch, which tracks hot groups and their write fractions in bounded memory, and per-tenant pacing for tenants that share the CPU and the stacks (\secref{subsection:design-sketch}, \secref{subsection:design-tenant});
  \item an evaluation of service capacity across five services, three NDP regimes, shifting workloads, and co-located services (\secref{sec:eval}).
\end{itemize}

\section{Background and Related Work}
\label{section:background}
\subsection{Service Capacity in Online Services}
\label{subsection:bg-online}

Online datacenter services, such as key-value stores~\cite{mao2012masstree,nishtala2013scaling}, transaction processing~\cite{tu2013silo}, search engines~\cite{xapian}, and recommendation systems~\cite{gupta2020recommendation}, must continuously serve incoming requests while meeting stringent tail-latency service-level objectives (SLO)~\cite{dean2013tail,barroso2018datacenter}.
A key problem in resource management of datacenter services is matching demand to capacity.
Demand comes from incoming requests and the work required to process them, while capacity is limited by the server's compute, memory, and bandwidth resources.
Most prior NDP work targets batch processing or throughput-oriented workloads~\cite{ahn2015tesseract,gomez2022benchmarking,chen2022pimcloud}, with performance typically measured by speedup or aggregate throughput.
Online services differ from throughput-oriented workloads in two important ways:
\romenum{1}~requests arrive individually, each involving a relatively small amount of work, and
\romenum{2}~request arrival times and data accesses are not known in advance.
Unlike batch processing, where work can be partitioned and scheduled ahead of time, online services must quickly make scheduling decisions as requests arrive~\cite{goder2015bistro,zhang2023shepherd}.
The performance metric also changes accordingly.
What matters more than throughput is the maximum request load a system can sustain while meeting its SLO, called \textbf{\textit{service capacity}}.
Service capacity and throughput might differ substantially.
Under heavy load, a system may maintain high throughput by letting many requests exceed the latency target~\cite{kaffes2019shinjuku}.
Conversely, allocating resources for worst-case SLOs can lead to substantial underutilization~\cite{lo2015heracles,ousterhout2019shenango}.
The goal of resource management is therefore to maximize the demand a system can serve within its SLO on given hardware.

\subsection{Why Demand Is Hard to Match}
\label{subsection:bg-demand}
Three characteristics of online-service workloads make it difficult to match demand to capacity.
\romenum{1} \emph{Demand changes over time.}
Request rates exhibit diurnal patterns, for instance in production key-value stores~\cite{atikoglu2012workload},
and the read/write mix differs widely across workloads:
of 153 Twitter cache clusters, most are read-dominant, yet more than 35\% are write-heavy and more than 20\% serve more writes than reads~\cite{yang2020twitter,yang2021large}.
Co-located services also compete for shared cores, cache capacity and memory bandwidth,
and the contention for these resources changes over time~\cite{chen2019parties,fried2020caladan,lin2026atumai}.
\romenum{2} \emph{Demand is unevenly distributed.}
Key popularity often follows a heavy-tailed distribution, with a small number of hot keys accounting for a large fraction of requests~\cite{atikoglu2012workload,cooper2010benchmarking}.
\romenum{3} \emph{Demand costs depend on data management.}
Each request must access specific data to complete.
Where a request executes therefore determines how it accesses that data and what synchronization is required.
Its execution cost depends on where it runs, where its data reside, and which other requests access the same data at the same time.
The load on a resource thus depends on the work that requests generate through their data accesses, beyond their number.
Matching demand to capacity requires knowing how much work is associated with each hot data item, where that work is performed, and how these patterns change over time.

\subsection{CPU+NDP Servers}
\label{subsection:bg-ndp}
A \cpundp server links a host CPU with several NDP stacks over an interconnect (\tabref{tab:system-config}).
The host CPU has several powerful out-of-order cores with private L1 and L2 caches and a shared last-level cache.
Each NDP stack is a memory stack, such as an HBM or HMC cube~\cite{jeddeloh2012hybriddram,lee2014hbm} or a CXL memory device~\cite{m2ndp}, whose logic layer holds a few small NDP cores.
Each NDP core has a private L1, and the cores of a stack share an L2 and reach the stack's local memory at high bandwidth and low latency.
Data are stored across the stacks, and each \emph{data item}, the smallest unit of data that a request reads or writes as a whole, has a \emph{home} stack.
A core can also reach data on other stacks across the interconnect, at a higher cost.

Requests enter at the host CPU, where a front end orders the requests of different tenants by their shares, and router threads then dispatch each request to an execution site, the CPU or an NDP stack, where a \emph{worker}, a core that executes requests, serves it.
On the CPU, the same host cores run the routers and the workers, so the CPU both dispatches every request and executes some of them.

This organization partitions capacity by data.
Demand, however, is skewed toward popular data and shifts over time, so a few stacks saturate while the others and the CPU have spare capacity.
Borrowing that capacity is practical, since any core can reach any data across the interconnect, but a request run away from its home stack reaches its data remotely, and data written on one side must be fetched again by the other (\secref{subsection:bg-observations}).

\myparagraph{The Role of the Host CPU}
As an execution site, the CPU differs from an NDP stack in two ways.
First, its capabilities complement those of the stacks: NDP cores have high-bandwidth access to local data but small caches, whereas the CPU's powerful cores and large caches suit hot, frequently reused data.
Second, its capacity is not tied to any data partition: as hot spots shift among the stacks, the CPU can lend its spare capacity to whichever partition is currently hot.
Because the CPU also dispatches every request, how its cores are split between dispatch and execution must follow the workload: read-heavy, fine-grained requests stress dispatch, while hot spots on the stacks call for more execution on the CPU.

\myparagraph{System Cache Coherence}
Our simulator adopts and implements hardware cache coherence across CPU and NDP caches, following prior NDP designs~\cite{ahn2015piminstructions,boroumand2019conda} and recent work on CXL~3.0-based near-data processing~\cite{cxl3,m2ndp}.
Within an NDP stack, cores maintain coherence through the shared L2.
Across CPU and NDP domains, a write invalidates cached copies in other domains, so subsequent reads must fetch the line again from its home stack.

Although hardware coherence is increasingly supported, software can provide the same guarantee on platforms that lack it: each request invalidates the group's cached lines after acquiring the group lock and writes back modified lines before releasing it.
Such lock-protected data are migratory, a sharing pattern for which software coherence performs comparably to hardware~\cite{adve1991comparison}.

\begin{table}[t]
  \centering
  \caption{Simulated server.}
  \label{tab:system-config}
  \vspace{-0.15in}
  \scriptsize
  \setlength{\tabcolsep}{3pt}
  \begin{tabularx}{\columnwidth}{@{}l>{\raggedright\arraybackslash}X@{}}
    \toprule
    Component & Configuration \\
    \midrule
    Host CPU & 2.4\,GHz out-of-order cores (Haswell-class): 24 CPU workers and 20 routing threads; pacing on separate cores \\
    CPU caches & 32\,KB L1, 512\,KB private L2, 64\,MB shared L3 \\
    NDP & 16 stacks $\times$ 8 small cores (Silvermont-class) at 2.0\,GHz \\
    NDP caches & 32\,KB L1 per core; 512\,KB L2 shared per stack \\
    Memory & 8 vault controllers per stack; HMC timing calibrated against Ramulator; closed page \\
    Links & 160\,GB/s between stacks; two CPU ports, 320\,GB/s in total \\
    \bottomrule
  \end{tabularx}
  \vspace{-0.25in}
\end{table}

\subsection{Resource Management for NDP Systems}
\label{subsection:bg-prior}
Two lines of prior work manage demand and capacity on NDP systems.
The first allocates NDP resources among cloud tenants, primarily at the granularity of virtual machines.
vPIM~\cite{teguia2024vpim} virtualizes UPMEM-NDP devices~\cite{upmem} so that several VMs share them in isolation.
PIMCloud~\cite{chen2022pimcloud}, the most related, is the first QoS-aware resource manager for cloud VMs with NDP.
Building on PARTIES~\cite{chen2019parties}, it co-locates latency-critical and best-effort applications and allocates CPU and NDP cores, along with their associated memory pages, to each application.
However, its core allocation decisions follow application-level resource preferences and QoS feedback, without explicitly deciding which hot-data requests the CPU should serve within an application.

The second decides where computation runs and where data lives.
PEI~\cite{ahn2015piminstructions} runs each PIM operation on the host or in memory according to measured data locality;
TOM~\cite{hsieh2016tom} offloads GPU code to the stack that holds its data and maps pages accordingly;
CoNDA~\cite{boroumand2019conda} keeps the host and near-data accelerators coherent at low cost.
ABNDP~\cite{tian2023abndp} jointly considers request placement and data placement.
Its scheduler assigns each task to the NDP unit that minimizes a weighted sum of remote-access cost and load imbalance.
Its \textit{Traveller Cache} stores cache-line copies of remote data at a few camp locations near the units that access them.
ABNDP targets batch workloads such as graph analytics, SpMV, and k-means, and evaluates speedup and energy consumption.
Tasks follow a bulk-synchronous model in which data read during a timestamp remain unchanged.
The cache therefore holds only read-only data.
Writes bypass it and go to the home location, and all cached copies are invalidated at the end of each timestamp.
All tasks execute on NDP units, and the host CPU is evaluated only as a separate CPU-only baseline.
Online services do not generally provide the phase boundaries that make this caching safe or the batch structure its scheduler assumes.

Online services on CPU+NDP servers require a combination that \textit{neither line of systems provides}: SLO-aware decisions made as requests arrive, at the granularity of individual hot data items, with the host CPU participating fully in request execution.
The CPU and NDP stacks must be able to serve the same hot data, and each decision must account for the coherence cost of that cooperation.

\section{Challenges and Motivating Observations}
\label{subsection:bg-observations}
\begin{figure*}[t]
  \centering
  \includegraphics[width=\textwidth]{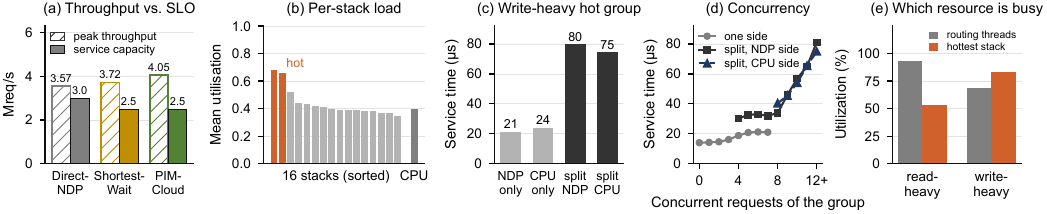}
  \vspace{-0.25in}
  \caption{\textbf{Demand and capacity on a CPU+NDP server.} (a)~Peak throughput and service capacity of three policies and (b)~utilization of each NDP stack and of the CPU under Shortest-Wait at 3.0\,Mreq/s, with 90\% writes and one hot-set move. (c)~Service time of a write-heavy hot group on its home stack, on the CPU, or split between them. (d)~Service time vs.\ concurrent requests of the group, on one side or split between the CPU and its stack (NDP and CPU side). (e)~Utilization of the routing threads and of the hottest stack's cores, under read- and write-heavy loads.}
  \label{fig:motivation}
  \vspace{-0.1in}
\end{figure*}

To illustrate the challenges of matching demand to capacity on a \cpundp server, we characterize a Masstree~\cite{mao2012masstree} key-value service in ZSim~\cite{sanchez2013zsim}.
Each request reads/writes a group of keys, with key popularity following a Zipf distribution~\cite{zipf2016human} with $\alpha=0.99$,
close to the median $\alpha$ reported in Twitter's production cache workloads~\cite{yang2020twitter}.
In the runs of \figref{fig:motivation}(a) and (b), the hot set shifts once: in production key-value stores, some key ranges are accessed intensively for only a short period~\cite{cao2020rocksdb}, and hot keys often cause load spikes~\cite{yang2020twitter}.
Each run measures 40{,}000 requests using the system configuration in \tabref{tab:system-config}, with a latency SLO of 99$\mu$s.
We compare three policies: Direct-NDP runs every request at its owner stack; Shortest-Wait sends each request to its owner stack or the CPU, whichever has the shorter estimated wait; and PIMCloud~\cite{chen2022pimcloud} routes requests preferably to the CPU, the side its offline profiling selects, and falls back to the owner stack when no CPU worker is free.

\myparagraph{Challenge 1: Peak Throughput $\neq$ Service Capacity}
In \figref{fig:motivation}(a), every policy completes more requests at its peak than it can serve within the SLO.
At an offered 4.0\,Mreq/s, for instance, Direct-NDP's throughput is 3.45\,Mreq/s, but 13\% of completed requests exceed the latency target; its service capacity is 3.0\,Mreq/s.
Across the three policies, peak throughput is 1.2--1.6$\times$ of service capacity.
Their ranking can also reverse: PIMCloud and Shortest-Wait reach 4--13\% higher peak throughput than Direct-NDP, yet sustain 17\% less load within the SLO (2.5 vs.\ 3.0\,Mreq/s).
Systems for online services must therefore be designed and evaluated by the load they can sustain under the SLO.

\myparagraph{Challenge 2: Skewed Demand Overloads a Few Partitions while Capacity is Underutilized}
In \figref{fig:motivation}(b), 2.4\% of requests miss the SLO, even though the median stack utilization is only 39\% and host CPU utilization is 40\%.
These SLO misses result from the two stacks that hold the hot data during the run.
Spare capacity remains idle on other stacks while the hot partitions are overloaded.

\myparagraph{Challenge 3: Sharing Write-Heavy Hot Data Must Be Bounded}
Borrowing spare capacity elsewhere changes the execution cost of a request, depending on how it accesses data.
\figref{fig:motivation}(c) shows a write-heavy hot group whose requests take 21\,$\mu$s each on the home stack and 24\,$\mu$s when served exclusively by the CPU, but 75--80\,$\mu$s on both sides when its requests are split between the two.
Two effects combine.
First, the requests of a write-heavy group synchronize on the same locks and data, so their service time grows with how many of them run at once (\figref{fig:motivation}(d)).
A stack alone caps this number at its own cores: on one side, service time rises from 14 to 21\,$\mu$s as up to seven other requests of the group run alongside.
Splitting the group between the stack and the CPU lets far more of its requests run at once, and beyond twelve each request takes 75--81\,$\mu$s, 5--6$\times$ its time at low concurrency.
Second, synchronization costs more across the two sides than within one.
Cores of one stack pass a lock or a written line through the stack's shared L2, but the CPU and an NDP stack share no cache, so a line written by one side must be fetched again from DRAM by the other.
At the same concurrency of five to eight, a request of a split group takes 32--34\,$\mu$s on the NDP side, against 21\,$\mu$s on one side.
This cost is specific to \cpundp servers: multicore CPUs synchronize through a shared last-level cache, and distributed-memory systems share no cache lines across servers.
A write-heavy group should therefore run on one side at a time, with a bounded number of concurrent requests.

\myparagraph{Challenge 4: A Hot Stack Near Capacity can Lock into Saturation}
The same contention makes a stack near capacity unstable.
On one side, the service time of a write-heavy group rises from 14 to 21\,$\mu$s as its requests in flight on the stack grow from two to five (\figref{fig:motivation}(d)).
The stack therefore has two stable states at the same offered load.
In the healthy state, queues stay short and each request takes about 14\,$\mu$s.
In the saturated state, many requests are in flight at once, each takes about 21\,$\mu$s, and the slower service keeps the queue long.
In our runs, a single poorly timed action, such as moving the group to another stack just as its write fraction rises,
pushes the stack into the saturated state, where about 6\% of all requests miss the SLO for the rest of the run.
Control actions must therefore avoid tipping a stack over this edge, even if that leaves some imbalance in place.

\myparagraph{Challenge 5: The System Bottleneck Shifts with the Workload}
Which resource limits service capacity depends on the workload (\figref{fig:motivation}(e)).
Under a read-heavy load at 5.5\,Mreq/s, the routing threads are 93\% busy while the most loaded stack is 53\% busy.
Under a write-heavy load at 4.0\,Mreq/s, the most loaded stack is 83\% busy and saturates first, while the routing threads still have spare capacity.
A fixed split of host cores between routing and execution, or a fixed placement of data, is therefore wrong for part of the workload,
and the system must rebalance as the workload shifts.

\myparagraph{Implications}
Together, these observations call for a management system that
(i)~targets service capacity under the SLO;
(ii)~decides, for each hot group, where its requests run and which site owns it, based on how the group is read and written;
(iii)~keeps each write-heavy group on one side with a bounded number of concurrent requests, and avoids actions that tip a stack into saturation; and
(iv)~follows the bottleneck as it moves between the front end, the CPU and the NDP stacks.
\secref{sec:design} presents such a system.

\section{\ourname Design}
\label{sec:design}

\subsection{Design Overview}
\label{subsection:design-overview}

We present \ourname, a management system for \cpundp servers that maximizes the service capacity
of latency-critical services by balancing load across the host CPU and the NDP stacks at the granularity of hot data.
An \emph{execution site} is either the host CPU or one of the NDP stacks.
\ourname tracks demand and manages data in units of \emph{groups}, sets of data items that requests often access together.
Each group has an \emph{owner}, the execution site that serves its requests by default.
Initially, the owner is the group's \emph{home} stack, the NDP stack that holds the group's data under the default placement; \ourname may later make another stack or the CPU the owner.
A few \emph{hot} groups, those with the highest request rates, carry much of the load;
\ourname manages their ownership individually, while the many other groups keep their home stack as owner.

\ourname makes two kinds of decisions: where each request executes, and how each group's data are managed.
Data management determines which site owns the group and serves its requests by default, which sites may cache the group's cache lines, and whether the group should migrate to its new owner.

Both decisions affect the \textit{normalized load} of each execution site $s$, defined as incoming service demand relative to execution capacity:
\[
  \rho_s \;=\; \frac{1}{n_s}\sum_{g} \lambda_{g,s}\, t_{g,s},
\]
where $\lambda_{g,s}$ is the rate of requests from group $g$ executed at $s$, $t_{g,s}$ is their service time there, and $n_s$ is the number of cores at $s$.
As $\rho_s$ approaches $1$, the queue at $s$ grows faster than it drains, and the waiting time of the requests it serves soon exceeds their SLO.
Request routing sets $\lambda_{g,s}$: sending requests from a hot stack to the CPU lowers the request rate at that stack.
Data management affects $t_{g,s}$: serving a request at its group's owner with local data avoids the cost of remote accesses and of coherence traffic from concurrent writes at another site.
The two decisions are coupled, because routing also changes $t_{g,s}$: splitting a write-heavy group's requests between the CPU and its stack makes both sites serve the group at once and lengthens service time on both sides (\figref{fig:motivation}(c)).
\ourname estimates these terms online to predict the effect of an action before taking it.

\begin{figure}[t]
  \centering
  \includegraphics[width=\columnwidth]{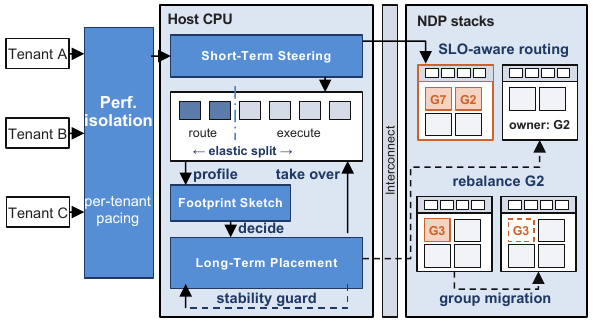}
  \vspace{-0.15in}
  \caption{\textbf{Overview of \ourname.} Dark boxes are its components; dashed lines are control actions. STS on the routing cores sends each request to its owner stack or a CPU worker. LTP uses the Footprint Sketch to rebalance hot groups and hand them to the CPU; its drift guards act only under drift. G2's ownership moves to a cold stack, and group migration moves G3's data to its owner's stack.}
  \label{fig:design-overview}
  \vspace{-0.12in}
\end{figure}

\figref{fig:design-overview} shows the structure of \ourname.
Requests enter through the front end, where per-tenant pacing (\secref{subsection:design-tenant}) limits each tenant's request rate to its share of the \cpundp server.
Routers on the host CPU then \textit{dispatch} each request to an execution site.
\ourname decides at two timescales.
\textbf{\textit{Short-Term Steering}} (STS, \secref{subsection:design-fast}) routes each request by predicting when it would complete at the group's owner and on the CPU:
a hot request stays at its owner if the owner can finish it within its remaining SLO budget, and any other request goes to the site that finishes first.
\textbf{\textit{Long-Term Placement}} (LTP, \secref{subsection:design-slow}) changes how hot groups are held, once per control period:
for each hot group on an overloaded stack, it picks a new owner, a lightly loaded stack or the CPU, and moves the entire group there.
Both rely on the \textbf{\textit{Footprint Sketch}} (\secref{subsection:design-sketch}), which samples requests to identify hot groups and estimate their write fractions.
LTP also has three \textbf{\textit{drift guards}} (\secref{subsection:design-slow}) that act only when a specific drift appears: a \textit{stability guard}, an \textit{elastic split} of host cores, and \textit{group migration}.

The two timescales reflect how quickly each decision takes effect.
Queues at a stack build up within tens of microseconds, so routing must follow current queue conditions.
A data-management action takes effect only after the group's in-flight requests at the old site finish or the group migrates.
LTP therefore acts once per control period $T_c$ (\tabref{tab:design-params}), based on the load and service times measured during that period.
STS absorbs short bursts under the current placement, and LTP corrects imbalance that persists across periods.

\subsection{Short-Term Steering: Routing Each Request}
\label{subsection:design-fast}
Routers on the host CPU run STS for each request as it arrives.
For a request $r$ of group $g$, the candidate sites are $g$'s owner $o(g)$ and the CPU, and the router predicts the finish time of $r$ at each candidate $x$:
\begin{equation}
  F_r(x) = \min_{k \in K_x} B_k \;+\; \delta \;+\; \hat{S}_x(h_g, b_r).
  \label{eq:finish}
\end{equation}
$\min_{k \in K_x} B_k$ is how long $r$ would wait at $x$, that is, at the core of $x$ that becomes free first.
For each core $k$, the router tracks $B_k$, the work still ahead of a newly arriving request.
$B_k$ adds up the predicted service times of the requests queued at $k$ and the remaining time of the request that $k$ is running.
$K_x$ is the set of cores of a stack, or $d$ CPU workers sampled at random, since scanning every CPU worker would exceed the router's time budget and a few samples balance nearly as well~\cite{mitzenmacher2001power}.
A worker qualifies only if sending $r$ to it keeps at most $C_{\max}$ requests of $g$ executing at once.
$\delta$ is the dispatch delay.
$\hat{S}_x(h_g, b_r)$ is the predicted service time of $r$ at $x$.
The router keeps one estimate per site, hot flag $h_g$ and write class $b_r$, which reflects the share of writes in the request.
Each estimate is a moving average of measured service times with weight $\alpha$.
The service times are measured where requests run.
The estimates therefore reflect the speed gap between CPU and NDP cores, remote accesses when a group runs away from its home stack, and coherence traffic when both sides run a group.

\begin{figure}[t]
  \centering
  \includegraphics[width=\columnwidth]{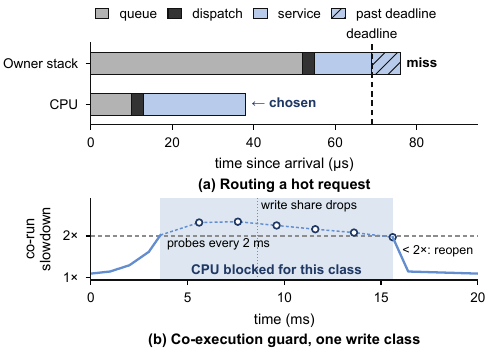}
  \vspace{-0.15in}
  \caption{\textbf{Short-Term Steering} (schematic). (a)~A hot request that would miss its deadline at the owner is sent to the CPU. (b)~The co-execution guard blocks the CPU for a write class once co-running slows it by $2\times$, and reopens it when probes every 2\,ms show the slowdown is gone.}
  \label{fig:design-route}
  \vspace{-0.15in}
\end{figure}

Each request has a deadline $d_r$: its SLO minus the time it has already spent in the system and a margin $m$.
The router compares the candidates in three steps.
\romenum{1} It prefers sites predicted to finish $r$ before $d_r$.
\romenum{2} Among these, a hot request stays at its owner, which keeps the hot data local and avoids co-execution.
\romenum{3} Otherwise, $r$ goes to the site with the earliest $F_r(x)$, so cold requests fill spare CPU capacity before the stacks build long queues, and a request no site can finish in time goes to the earliest one (\figref{fig:design-route}(a)).

Sending some of a hot group's requests to the CPU has a cost that no single request sees: both sides then run the group, and for a write-heavy group every request on both sides slows down (\figref{fig:motivation}(c)).
STS therefore stops offering the CPU in two cases.
First, the stability guard of LTP may pin a write-heavy group that runs near capacity to its home stack (Challenge~4, \secref{subsection:design-slow}), and STS then keeps all of its requests there.
Second, the co-execution guard watches for the slowdown directly.
For each write class, it divides the service time measured while both sides run the class by the service time measured while only one side does.
If this ratio reaches $\theta$ on either side, CPU no longer takes hot requests of that class; $\theta$ is set well above the 1.26--1.5$\times$ fluctuation we measure without co-execution, so noise does not trigger it.
A probe still goes to the CPU periodically, and the class reopens once the slowdown is gone (\figref{fig:design-route}(b)).

\subsection{Long-Term Placement: Managing Hot Data}
\label{subsection:design-slow}
STS relieves a hot stack one request at a time.
This works for read-heavy groups, whose requests can run on the CPU at little extra cost.
It works poorly for write-heavy groups: splitting their requests between the CPU and the stack slows both sides (\secref{subsection:bg-observations}), so the co-execution guard keeps these requests at their owner.
When several write-heavy hot groups share a stack, STS therefore cannot relieve it.
LTP instead moves entire groups to a new owner, so their requests no longer need to be split.
It has two actions.
\emph{Rebalance} makes a lightly loaded stack the group's owner, and \emph{Takeover} makes the CPU its owner.
Either way the write-heavy group keeps a single owner, so it never runs on two sites at once.

Workers record the service time of each request when it completes, separately for CPU and NDP execution.
In each control period, LTP uses these records to compute the utilization of each site $s$, $U_s = \sum_{q \in C_s} S_q / (n_s T_c)$, where $n_s$ is the number of worker cores at $s$, $C_s$ is the set of requests completed at $s$ in the period, and $S_q$ is their service time.
LTP also tracks $B_s$, the predicted service time of the requests waiting in the queue of $s$.
LTP acts on a stack only when two conditions hold.
First, the stack is busy: its utilization or its queue exceeds a threshold ($U_s \ge U_{\mathrm{hi}}$ or $B_s \ge B_{\mathrm{hi}}$).
Second, it is busier than the others: $U_s \ge \bar{U} + \Delta_U$, where $\bar{U}$ is the mean utilization of all stacks.
The second condition keeps LTP from moving groups back and forth among stacks that are equally busy.

Algorithm~\ref{alg:ltp} applies three actions in order: Rebalance, Takeover and Return.
Each action handles at most $M$ groups per period, starting with the group that adds the most load to its stack.
The algorithm uses the following terms for a group $g$ on stack $s$.
$a_g$ is the fraction of the worker time of $s$ that $g$ uses.
$c_g$ is the share of the busy time of $s$ that comes from $g$.
$D_g$ is the number of NDP cores that $g$ needs.
$S^{\mathrm{cpu}}_g$ and $S^{\mathrm{ndp}}_g$ are the service times of $g$ on the CPU and on its stack.
$\hat{U}_{\mathrm{cpu}}$ is the predicted CPU load if the CPU takes over $g$, and $G_{\mathrm{cpu}}$ is the set of groups that the CPU owns.
\tabref{tab:design-params} lists the parameter values.

\begin{algorithm}[t]
  \small
  \caption{Long-Term Placement, once per control period $T_c$}
  \label{alg:ltp}
  \SetKwFunction{Move}{Move}
  \SetKwFunction{CpuCanTake}{CpuCanTake}
  \SetKwProg{Fn}{function}{:}{}
  \SetKw{Continue}{continue}
  \SetKw{KwAnd}{and}
  $\mathcal{O} \gets \{\, s : (U_s \ge U_{\mathrm{hi}} \lor B_s \ge B_{\mathrm{hi}}) \land U_s \ge \bar{U} + \Delta_U \,\}$\;
  \tcp{Rebalance}
  \ForEach{tracked group $g$ owned by a stack $s \in \mathcal{O}$, busiest first, up to $M$}{
    \lIf{$g$ is pinned}{\Continue}
    $t \gets$ least loaded stack with $U_t < \bar{U}$\;
    \lIf{$U_t + \kappa\, a_g < \max(U_{\mathrm{hi}},\, U_s)$}{\Move{$g, t$}}
  }
  \tcp{Takeover}
  \ForEach{tracked group $g$ owned by a stack $s \in \mathcal{O}$, busiest first, up to $M$}{
    \lIf{$g$ is not pinned \KwAnd \CpuCanTake{$g$}}{\Move{$g, \mathrm{CPU}$}}
  }
  \tcp{Return}
  \ForEach{group $g$ owned away from its home $h$ for at least two periods, up to $M$}{
    \lIf{$U_h + a_g \le U_{\mathrm{lo}}$ \KwAnd arrivals have not just dropped briefly}{\Move{$g, h$}}
  }
  \Fn{\CpuCanTake{$g$}}{
    \Return $c_g \ge c_{\min}$ \KwAnd $D_g \le D_{\max}$ \KwAnd $S^{\mathrm{cpu}}_g \le \sigma S^{\mathrm{ndp}}_g$ \KwAnd $\hat{U}_{\mathrm{cpu}} \le U^{\mathrm{cpu}}_{\mathrm{hi}}$ \KwAnd $|G_{\mathrm{cpu}}| < G_{\max}$\;
  }
  \Fn{\Move{$g, x$}}{
    mark $g$ moving; publish $x$ as its owner\;
    $x$ starts $g$ once the old owner has drained $g$, or after $T_{\mathrm{fence}}$\;
  }
\end{algorithm}

\begin{table}[t]
  \centering
  \caption{Parameters of \ourname.}
  \label{tab:design-params}
  \vspace{-0.15in}
  \scriptsize
  \setlength{\tabcolsep}{3pt}
  \begin{tabularx}{\columnwidth}{@{}>{\raggedright\arraybackslash}X>{\raggedright\arraybackslash}p{.36\columnwidth}@{}}
    \toprule
    Parameter & Value \\
    \midrule
    STS: dispatch delay $\delta$; deadline margin $m$; smoothing $\alpha$; CPU cores sampled $d$; concurrency cap $C_{\max}$ & 3\,$\mu$s; 30\,$\mu$s; $2^{-3}$; 4; 8 \\
    Co-execution guard: threshold $\theta$; window; probe & $2\times$; 64 samples; every 2\,ms \\
    Control period $T_c$; drain timeout $T_{\mathrm{fence}}$ & 250\,$\mu$s; 10\,ms \\
    Overload: $U_{\mathrm{hi}}$; $B_{\mathrm{hi}}$; $\Delta_U$ & 60\%; 20\,$\mu$s; 15 points \\
    Return home at $U_{\mathrm{lo}}$; remote cost $\kappa$ & 50\%; 1.6 \\
    Takeover: $c_{\min}$; $D_{\max}$; $\sigma$; $U^{\mathrm{cpu}}_{\mathrm{hi}}$; $G_{\max}$ & 2\%; $C_{\max}/2$; 1.2; 75\%; 16 \\
    Moves per action and period $M$ & 8 \\
    Stability guard: pin at write fraction $w_{\mathrm{pin}}$; unpin after & 50\%; 4 periods below 0.8$\times$ \\
    Mix-change pause: margin; average weight; length & 20 points; 1/4; 8 periods \\
    Elastic split: target $\rho^{*}$; period; min.\ threads per side; threads returned per period & 80\%; 50\,$\mu$s; 8; at most 8 \\
    Sketch: sampling; Space-Saving; tracked groups; aging & 1 in 4; 4096; 32; halve every 4 epochs of 250\,$\mu$s \\
    \bottomrule
  \end{tabularx}
  \vspace{-0.25in}
\end{table}

\emph{Rebalance} picks the least loaded stack among those below the mean utilization and makes it the group's owner.
The group's data stay at its home stack until group migration moves them (\secref{subsection:design-slow}).
Until then, requests at the new owner read the data remotely and run slower.
LTP thus counts the group as $\kappa\,a_g$ of load at the new owner, where $\kappa$ is the measured slowdown of remote access.
It makes the move only if the new owner's load then stays below $\max(U_{\mathrm{hi}}, U_s)$.
In other words, the new owner must stay below the busy threshold, or at least below the old owner's prior load.

\emph{Takeover} makes the CPU the owner of an entire hot group, so the CPU serves all of its requests.
LTP takes over a group only if four conditions hold.
The group is large enough to matter ($c_g \ge c_{\min}$) and small enough for the CPU ($D_g \le D_{\max}$).
It is not much slower on the CPU than on its stack ($S^{\mathrm{cpu}}_g \le \sigma S^{\mathrm{ndp}}_g$).
The predicted CPU load stays at most $U^{\mathrm{cpu}}_{\mathrm{hi}}$.
The CPU owns at most $G_{\max}$ groups.

\emph{Return} moves a group back home after at least two periods away, once its home load plus $a_g$ is at most $U_{\mathrm{lo}}$ (below $U_{\mathrm{hi}}$).
A short drop in arrivals does not trigger a return, since it says little about the next burst.
All three actions skip groups pinned by the stability guard (\secref{subsection:design-slow}).

The limit $D_{\max}$ matters most for write-heavy groups.
STS lets at most $C_{\max}$ requests of a group run at once on either side.
On its owner stack, these requests pass each lock through the stack's shared L2.
On the CPU, each lock transfer crosses the CPU--NDP link, so the same number of requests runs slower (\secref{subsection:bg-observations}).
$D_{\max}$ bounds the group's average demand, while $C_{\max}$ bounds how many of its requests run at the same time.
We set $D_{\max} = C_{\max}/2$, so that a group on the CPU rarely reaches the cap and its lock rarely becomes a bottleneck.
Keeping such a group on one stack follows the idea of lock delegation~\cite{lozi2012rcl,roghanchi2017ffwd}, which runs a critical section on one dedicated core instead of letting many cores compete for its lock.

Every change of owner waits for the old owner to finish.
Requests already sent to the old owner complete there.
The new owner starts the group's requests only after the old owner has none left, or after a timeout $T_{\mathrm{fence}}$.
The group therefore never runs on both sides during a move, so its requests cause no coherence traffic between the two sides and never race on its data.

LTP also decides which sites may cache a group's lines, based on its write fraction.
For a write-heavy group, with at least 70\% writes, only the owner and the CPU may cache its lines.
A write then never has to invalidate copies spread over several stacks.
A read-heavy group, with at most 30\% writes, may be cached wherever it runs.

\myparagraph{Drift guards}
The actions above run in every period.
LTP also has three \emph{drift guards}.
Each stays idle while the workload is steady and acts only when one kind of drift appears.

The first guard, the \emph{stability guard}, protects a stack that runs near capacity.
On such a stack, one badly timed action can push the stack into the saturated state of Challenge~4.
Examples are moving a write-heavy hot group or splitting its requests.
To find which actions are safe there, we replayed four runs that ended in this state, each with the hot group handled by one fixed policy.
Keeping the group entirely at its home stack missed at most 0.61\% of requests in every run.
Freezing all moves came close, at most 0.98\%, but still let STS split the group's requests.
Moving the group to the CPU missed up to 10.2\%, and sending half of its requests to the CPU missed 4.9--8.3\%.
The guard therefore pins a group to its home stack when two conditions hold: the group is write-heavy ($w_g \ge w_{\mathrm{pin}}$), and it needs more than $D_{\max}$ cores.
STS then offers a pinned group no other site, and LTP does not move it.
The group is unpinned once its write fraction or its demand has stayed below these thresholds, by a margin, for several periods.

The stability guard also pauses LTP while the read/write mix of the workload changes.
The sketch tracks the write fraction over all sampled requests and a slowly moving average of it.
When the two differ by more than a margin, the service times measured so far no longer describe the workload.
The guard then stops all moves for a few periods and resets the per-group service estimates.
STS keeps routing every request during this time and uses spare capacity to absorb the change.

The second guard, the \emph{elastic split}, moves host cores between routing and execution as the bottleneck shifts between them (\figref{fig:motivation}(e)).
Read-heavy loads at high rates leave the CPU workers idle while requests queue at the routers.
Write-heavy loads whose hot groups move to the CPU keep the workers busy while the routers have spare capacity.
The guard treats the host threads, one per core, as one pool in which each thread either routes or executes.
In every short period, it measures the arrival rate $\lambda$ and the mean routing time $t_r$.
It then sizes the routing side for a target utilization $\rho^{*}$:
\begin{equation}
  R \;=\; \Bigl\lceil \max\bigl(\lambda\, t_r,\; \rho_R R_{\mathrm{now}}\bigr) / \rho^{*} \Bigr\rceil ,
  \label{eq:elastic}
\end{equation}
where $\rho_R$ is the current utilization of the $R_{\mathrm{now}}$ routing threads.
It keeps a minimum number of threads on each side.
It adds routing threads at once, because a routing queue grows quickly, and returns threads to execution a few at a time.
Before a thread starts routing, the routers stop sending it requests and it finishes the requests already in its queue, so no request is lost.

The third guard, \emph{group migration}, moves a group's data to its owner.
A group's owner can differ from the stack that holds its data, because of the initial placement or because LTP moved the owner.
Its requests then access the data remotely.
The guard copies the data of such a hot group to the owner stack and switches their home once the copy is complete.
Data that several groups share stay where they are.
Data that the group writes later are allocated at the owner stack, keeping new data local as well.

\subsection{Footprint Sketch}
\label{subsection:design-sketch}
\begin{figure}[t]
  \centering
  \includegraphics[width=\columnwidth]{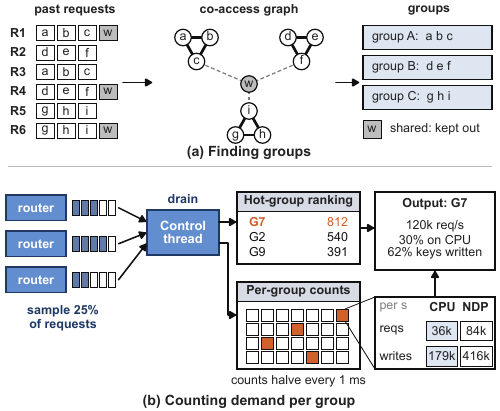}
  \vspace{-0.25in}
  \caption{\textbf{The Footprint Sketch.} (a)~Items often accessed together are merged into groups with union-find; an item that links several groups (w) stays out of all of them. (b)~Routers sample one request in four. A Space-Saving summary ranks hot groups, and a Count-Min sketch counts each group's requests and writes on the CPU and NDP sides; counters halve every millisecond.}
  \label{fig:design-sketch}
  \vspace{-0.15in}
\end{figure}

STS and LTP need to know the hot groups, and LTP also needs the write fraction of each hot group.
Exact per-group counts are not practical.
A service can hold millions of groups, the hot set changes within milliseconds, and an exact table would add a shared update to the dispatch of every request.
The Footprint Sketch therefore estimates these statistics from sampled requests in fixed memory.
In our runs, the 32 groups it tracks include 81--91\% of the true 32 hottest.
Service times are measured only for these tracked groups.

Requests of online services often access data in recurring combinations.
Examples are the keys of a multi-key request and the district, customer and order rows of a TPC-C transaction~\cite{tpcc}.
Database partitioners build on the same observation~\cite{curino2010schism}.
The sketch links two items when enough sampled requests access both, and it merges linked items into groups with union-find~\cite{tarjan1975union} (\figref{fig:design-sketch}(a)).
An item shared by several groups, such as a TPC-C warehouse row, stays out of all groups and remains at its home stack.
Each group is counted, assigned an owner and moved as a unit.
Its items may span several stacks, and the remote accesses this causes show up in the group's measured service time.
Groups are much smaller than a stack's data partition, so \ourname can manage the few groups behind a hot spot individually.

During dispatch, each router samples one request in four and appends a record to its own ring buffer.
The record holds the group, the number of items the request writes, its execution side and its home stack.
Sampling therefore needs no locks or shared counters on the dispatch path.
A control thread drains the rings periodically and updates two structures (\figref{fig:design-sketch}(b)).
A Space-Saving summary~\cite{metwally2005spacesaving} with 4096 counters ranks the groups by request count, and the controller tracks the top 32 in each control period.
A Count-Min sketch~\cite{cormode2005countmin} with four rows of 4096 cells records how each group is accessed.
Each cell holds four counters: requests sent to the CPU, requests sent to the NDP stacks, and items written on each side.
For a group $g$, each count is the minimum across the four rows.
From these counts, the controller derives the request rate $\lambda_g$ and the share of requests on the CPU.
It also derives the write fraction $w_g$, the number of items written over the number of items accessed.
Space-Saving tells which groups are hot, and Count-Min gives their statistics, also for groups that have just entered the top.
The control thread also tracks the write fraction over all samples to detect shifts in the read/write mix (\secref{subsection:design-slow}).

All counters are halved every few epochs, as in the reset of TinyLFU~\cite{einziger2017tinylfu}.
A group that stops receiving requests thus loses half its count at each halving, and newly hot groups soon overtake it.

\subsection{Performance Isolation}
\label{subsection:design-tenant}

When several tenants share the server, a burst from one tenant fills the front end and the shared queues.
The other tenants' requests then miss their SLO even though their own load has not changed.
\ourname isolates tenants by giving each one a share of the server's capacity and enforcing the shares where requests enter, before the routers.
Like the drift guards, this mechanism is idle until a tenant exceeds its share.
Each tenant $u$ has a token bucket $\tau_u$ with rate $r_u$, its share of the server's capacity, refilled every $\Delta$:
\begin{equation}
  \tau_u \;\leftarrow\; \min\bigl(r_u T,\; \tau_u + r_u \Delta\bigr),
  \label{eq:bucket}
\end{equation}
where the depth $T$ bounds the burst a tenant can send at once.
A global bucket, refilled the same way, limits the total forwarded rate to the capacity of the server.
A request within its tenant's share is forwarded at once.
A request above the share waits in a per-tenant queue until its tenant's bucket refills.
If the front end has spare capacity, the request may instead borrow a token from the global bucket.
When the global bucket is empty as well, the request keeps waiting in its tenant's queue.
A request that waits past its SLO counts as a miss.
A single pacing thread refills the buckets, and tokens are taken with atomic updates, so pacing adds no lock to the request path.
A tenant's burst thus consumes its own share and the server's spare capacity, leaving other tenants' shares intact.
    
\section{Evaluation}
\label{sec:eval}

\providecommand{\evaltakeaway}[1]{%
  \par\smallskip
  \begingroup
  \setlength{\fboxsep}{4pt}%
  \noindent\colorbox{black!5}{%
    \parbox{\dimexpr\columnwidth-2\fboxsep\relax}{%
      \textbf{Takeaway.} #1}}%
  \endgroup
  \par\smallskip}

\subsection{Experimental Methodology and Setup}
\label{subsection:nv-method}

We evaluate \ourname in ZSim~\cite{sanchez2013zsim}, with memory timing calibrated against Ramulator~\cite{kim2016ramulator}.
ZSim tracks the arrival time, execution site, and memory and interconnect traffic of each request, so end-to-end latency includes the queueing effects of routing and data placement.
Unless otherwise stated, all methods use the same hardware configuration (\tabref{tab:system-config}), request arrivals and measurement window.

\begin{table}[t]
  \centering
  \caption{Evaluated services.}
  \label{tab:nv-workloads}
  \vspace{-0.15in}
  \scriptsize
  \setlength{\tabcolsep}{3pt}
  \begin{tabularx}{\columnwidth}{@{}lll>{\raggedright\arraybackslash}X@{}}
    \toprule
    Case & Engine & Workload & Description \\
    \midrule
    U1 & Masstree~\cite{mao2012masstree} & YCSB~\cite{cooper2010benchmarking} & Hot data local to one stack; write share and hot set drift \\
    U2 & Masstree & YCSB & Every request touches several stacks \\
    U3 & Masstree & YCSB & Read-heavy and write-heavy hot groups side by side \\
    U4 & Silo~\cite{tu2013silo} & TPC-C~\cite{tpcc} & Hot warehouses; four transaction types \\
    U5 & Redis~\cite{redis} & Pipelined & Multi-key GET, SET, MGET, MSET, INCR \\
    Others & TailBench~\cite{kasture2016tailbench} & 4 apps & Xapian~\cite{xapian} (search), Moses~\cite{koehn2007moses} (translation), Sphinx~\cite{huggins2006pocketsphinx} (speech), Img-DNN~\cite{yuan_multilayer_convnet} (image) \\
    \bottomrule
  \end{tabularx}
  \vspace{-0.1in}
\end{table}

\begin{table}[t]
  \centering
  \caption{Evaluated systems.}
  \label{tab:nv-systems}
  \vspace{-0.15in}
  \scriptsize
  \setlength{\tabcolsep}{3pt}
  \begin{tabularx}{\columnwidth}{@{}l>{\raggedright\arraybackslash}X@{}}
    \toprule
    System & Routing and placement \\
    \midrule
    CPU-only & Original benchmark engine implementation on the CPU, without NDP \\
    Direct-NDP & All requests at the owner stack \\
    PIMCloud~\cite{chen2022pimcloud} & Offline profiling picks the preferred side: Idle CPU first, else the owner stack; cautious replication of hot read-only pages \\
    ABNDP~\cite{tian2023abndp} & Load- and locality-aware routing over the stacks and the CPU; Traveller Cache of remote data in stack DRAM \\
    \addlinespace[\aboverulesep]
    \hdashline
    \addlinespace[\belowrulesep]
    Shortest-Wait (ablation) & \ourname variant: Owner stack or CPU, whichever has the shorter wait \\
    \rowcolor[gray]{0.94}
    \ourname & STS and LTP \\
    \bottomrule
  \end{tabularx}
  \vspace{-0.1in}
\end{table}

\tabref{tab:nv-workloads} summarizes the service workloads.
\tabref{tab:nv-systems} lists \ourname, its ablation variants and the baseline systems.
PIMCloud and ABNDP~\cite{chen2022pimcloud,tian2023abndp} follow the policies of their original designs, and we run them on the same services, request arrivals and hardware as \ourname.
The three Masstree services differ in data locality and read/write mix, the two properties that decide whether a request should run on the CPU or on NDP.
PIMCloud, ABNDP and \ourname execute requests on both the CPU and the NDP stacks.
ABNDP was designed for NDP-only batch workloads, so we add the CPU as one more site its scheduler can choose.
All systems that run requests on both sides cache NDP data on the CPU coherently.
Our main metric is \emph{service capacity}: the highest offered load at which at most 1\% of requests miss the SLO.
We also report SLO misses over time (\secref{subsection:nv-timeline}), per-tenant SLO misses (\secref{subsection:nv-isolation}) and p99 latency for the TailBench services (\secref{subsection:nv-tailbench}).
Each workload's SLO is about $5\times$ its mean service time.

\subsection{Main Results in Service Capacity}
\label{subsection:nv-main}
\begin{figure}[t]
  \centering
  \includegraphics[width=\columnwidth]{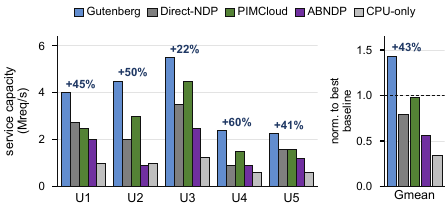}
  \vspace{-0.15in}
  \caption{\textbf{Service capacity.} Labels give \ourname's gain over the best baseline of each service; Gmean is normalized to that baseline. Services as in \tabref{tab:nv-workloads}.}
  \label{fig:nv-main}
  \vspace{-0.15in}
\end{figure}

\figref{fig:nv-main} shows that \ourname improves service capacity over the best baseline by 45\% on U1, 50\% on U2, 22\% on U3, 60\% on U4 and 41\% on U5, and by 43\% on geometric mean.
PIMCloud is the strongest baseline on U2, U3 and U4.
It prefers the CPU and falls back to the owner stack once the CPU workers are busy.
As load rises, most requests of a hot group therefore queue at its NDP stack.
Direct-NDP uses the CPU only for dispatch, so its capacity is bounded by the most loaded stack.
ABNDP balances load across NDP units and caches remote data in its Traveller Cache.
Its routing counts the cost of CPU accesses over the CPU--NDP link, so hot requests tend to stay on the NDP side near their data.
CPU-only accesses all data over the CPU--NDP link, reaching at most 1.25\,Mreq/s.

\evaltakeaway{Across five services, \ourname raises service capacity by 22--60\% (43\% on geometric mean) over the best baseline. SLO-aware routing balances each request between the CPU and the NDP stacks. Placement of hot data balances load among the NDP stacks and keeps each write-heavy group on a single site. The capacity of the CPU and the NDP stacks therefore adds up.}

\subsection{Ablation Study: STS and LTP}
\label{subsection:nv-ablation}
\begin{table}[t]
  \centering
  \caption{Service capacity of \ourname relative to each ablation variant on Masstree key-value workloads.}
  \label{tab:nv-ablation}
  \vspace{-0.1in}
  \small
  \begin{tabular}{@{}lcc@{}}
    \toprule
    \ourname vs. & Geometric mean & Maximum \\
    \midrule
    Shortest-Wait (without STS) & 1.18$\times$ & 1.60$\times$ \\
    \ourname without LTP & 1.26$\times$ & 1.67$\times$ \\
    \bottomrule
  \end{tabular}
  \vspace{-0.15in}
\end{table}

We evaluate STS and LTP on Masstree key-value workloads that vary in three ways: the write share (90\%, or rising from 10\% to 90\% mid-run), how often the hot set moves (never, once or three times), and how the hot data are spread (one Zipf hot set, or four medium write-heavy groups on one stack).
We compare \ourname{} with two ablation variants:
\romenum{1} Shortest-Wait, which replaces STS with a policy that routes to the owner stack or the CPU, whichever has the shorter estimated wait,
and \romenum{2} \ourname{} without LTP, which fixes each group's owner at its home stack.
On geometric mean, \ourname{} reaches 1.18$\times$ the service capacity of Shortest-Wait and 1.26$\times$ that of \ourname{} without LTP, and up to 1.60$\times$ and 1.67$\times$ (\tabref{tab:nv-ablation}).

\myparagraph{What STS adds}
Shortest-Wait uses a single service-time estimate for each side and splits write-heavy hot groups between the CPU and their stack.
After a hot-set shift, both sides serve the newly hot group, and service times grow on both sides.
STS avoids this costly co-execution by keeping each affected group's requests at one execution site.
STS also relies on CPU-side caching.
Without it, every CPU access crosses the CPU--NDP link, and capacity on the Zipf workloads drops from 4.0 to 2.0--2.5\,Mreq/s.

\myparagraph{What LTP adds}
LTP matters most when several write-heavy groups share one stack.
No group alone overloads the stack, but together they saturate it.
Without LTP, \ourname{} keeps their write-heavy requests at the owner to avoid the slowdown of concurrent CPU execution.
LTP detects the overload and moves entire groups to the CPU or a lightly loaded stack.
With 90\% writes and one hot-set shift, \ourname{} reaches 5.0\,Mreq/s, against 3.0\,Mreq/s without LTP.
Without LTP, the normalized load of the newly hot stack reaches 1.5, and its queueing delay grows to 2.3\,ms.
With LTP, \ourname{} moves three groups to the CPU and one to a lightly loaded stack within 0.9\,ms of the shift, and the utilization of the hot stack drops from 0.92 to 0.63.

\evaltakeaway{The SLO-aware routing of STS balances load between the CPU and the NDP stacks without splitting write-heavy groups, and LTP adds capacity where several write-heavy groups load one stack, by moving entire groups.}

\subsection{Following a Shifting Workload}
\label{subsection:nv-timeline}

\begin{figure}[t]
  \centering
  \includegraphics[width=\columnwidth]{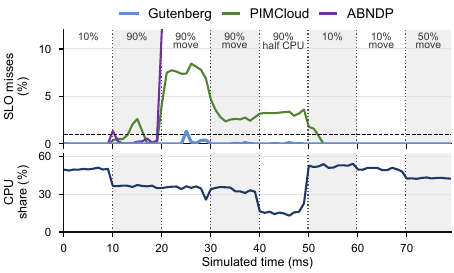}
  \vspace{-0.3in}
  \caption{\textbf{Following a shifting workload.} Top: SLO misses per 1\,ms window (dashed: 1\% budget). Bottom: share of requests \ourname runs on the CPU. Labels show the write share; \emph{move}: the hot set moves; \emph{half CPU}: half of CPU workers are taken. ABNDP leaves the plotted range at 20\,ms and does not recover.}
  \label{fig:nv-timeline}
  \vspace{-0.15in}
\end{figure}

Online workloads often change over time: the read/write mix shifts, the hot set moves, and co-located jobs may take CPU capacity.
We run the key-value service at 3.0\,Mreq/s through eight phases of about 10\,ms that combine these changes (\figref{fig:nv-timeline}):
the write share moves between 10\%, 50\% and 90\%, the hot set moves four times, and in one phase a co-located job takes half of the CPU workers.

Over the whole run, \ourname misses the SLO for 0.04\% of requests, against 1.85\% for PIMCloud.
ABNDP sustains only 2.0\,Mreq/s on this service; it falls behind once the hot set moves under 90\% writes and does not recover.
PIMCloud adapts how many cores serve the service, but its page actions only migrate private pages and replicate read-only ones~\cite{chen2022pimcloud}.
A writable hot group therefore stays where it is, and its requests either queue at its stack or are split with the CPU.
When the hot set moves under 90\% writes, PIMCloud's misses rise to about 8\%, and they stay above the 1\% budget until the workload turns read-heavy again.
\ourname misses at most 0.23\% in any phase.
The lower panel shows what \ourname does in each phase.
In read-heavy phases, the CPU executes about half of all requests.
When the workload turns write-heavy, the CPU share drops to about 35\%: requests of write-heavy hot groups stay with their owner,
and LTP moves entire hot groups to the CPU (3 times) or to cold stacks (9 times).
When the co-located job takes half of the CPU workers, the share falls to about 18\%, and the stacks absorb the rest.

\evaltakeaway{When the read/write mix, the hot set and the available CPU change over time, \ourname keeps SLO misses at 0.04\%. PIMCloud cannot move writable hot data and misses up to about 8\% in write-heavy phases.}

\subsection{Drift Guards}
\label{subsection:nv-guards}
\begin{table}[t]
  \centering
  \caption{SLO misses (\%) with each mechanism off and on, under the drift it targets and under a steady workload of the same service.\; $^{*}$: Between bursts, other tenants.}
  \label{tab:nv-guards}
  \vspace{-0.1in}
  \scriptsize
  \setlength{\tabcolsep}{3pt}
  \begin{tabularx}{\columnwidth}{@{}l>{\raggedright\arraybackslash}Xcc@{}}
    \toprule
    & & With the drift & Without it \\
    Mechanism & Drift it targets & off $\rightarrow$ on & off $\rightarrow$ on \\
    \midrule
    Stability guard & Read/write mix shifts & 1.76 $\rightarrow$ 0.01 & 0.16 $\rightarrow$ 0.39 \\
    Group migration & Data away from its owner & 4.67 $\rightarrow$ 0.28 & 0.26 $\rightarrow$ 0.26 \\
    Elastic split & Bottleneck moves to routing & 98.9 $\rightarrow$ 0 & 0.10 $\rightarrow$ 0.12 \\
    Co-execution guard & Hot set shifts & 1.54 $\rightarrow$ 0.71 & 0 $\rightarrow$ 0 \\
    Per-tenant pacing & A tenant bursts & 45.9 $\rightarrow$ 15.2 & 13.2 $\rightarrow$ 2.6$^{*}$ \\
    \bottomrule
  \end{tabularx}
  \\[2pt]
  \vspace{-0.15in}
\end{table}

\tabref{tab:nv-guards} runs each guard, off and on, in the drift it targets and in a workload without it.

\myparagraph{Stability guard}
The write share of the key-value service rises from 10\% to 90\% at 4.0\,Mreq/s.
A write-heavy hot group can then build a queue on its stack that never drains, and \ourname without the guard misses 1.76\% of requests.
With the guard, which pins the group at home, the misses drop to 0.01\%.
With a steady 90\% writes, where no such queue forms, both stay within the SLO.

\myparagraph{Group migration}
We place the data of the 64 hottest key-value groups on a stack other than their owner's.
Migration moves them to their owners within the first 10\% of the run and raises the capacity from 3.0 to 4.0\,Mreq/s, by 33\%.
The share of local accesses of these groups rises from 37\% to 62\%, and their service time on the NDP cores drops from 23.9 to 21.7\,$\mu$s.
The copies total about 1.7\,MB per run.
When the data are already on the right stack (U2), migration has nothing to do and the capacity is unchanged.

\myparagraph{Elastic split}
A read-heavy key-value load is limited by routing: at 7.0\,Mreq/s, the default 20 routing threads fall behind and 98.9\% of requests miss the SLO.
The guard grows the routing side to about 33 threads, and no request misses, matching the best fixed split for this load.
When execution is the limit, the guard keeps the split near its default.
In the shifting workload of \secref{subsection:nv-timeline}, it misses 0.12\% of requests against 0.10\% with the fixed split.
On the write-heavy loads of \secref{subsection:nv-ablation}, its capacity equals that of the fixed split.

\myparagraph{Co-execution guard}
The co-execution guard of STS (\secref{subsection:design-fast}) behaves like a drift guard.
When the hot set of U1 drifts, requests of a newly hot write-heavy group start to run on the CPU and on their stack at the same time.
Without the guard, such requests take 121--149\,$\mu$s, against 25--29\,$\mu$s when one side runs them alone, and the capacity falls from 4.0 to 3.5\,Mreq/s.
The guard detects the slowdown and keeps the group on one side, which brings these requests down to 33--56\,$\mu$s.
On a steady read-heavy load, it acts about 50 times per run, and both settings reach 5.5\,Mreq/s.

\evaltakeaway{Each guard restores the SLO in the drift it targets and changes little when that drift is absent.}

\subsection{Performance Isolation}
\label{subsection:nv-isolation}

Five tenants share the server, each modeled on a production cache workload:
three Facebook memcached pools (ETC, USR, VAR)~\cite{atikoglu2012workload} and two Twitter cache clusters~\cite{yang2020twitter}, one with hot keys and one write-heavy.
The tenants offer 3.5\,Mreq/s in total, and the write-heavy tenant periodically bursts to twice that rate (last row of \tabref{tab:nv-guards}).
Without per-tenant pacing, the other tenants miss their SLO for 46\% of their requests during the bursts, because their requests wait at the front end before they are routed.
Per-tenant pacing (\secref{subsection:design-tenant}) holds each tenant to its share at the front end.
The other tenants' misses then drop to 15\% during the bursts and from 13\% to 2.6\% between them, and the bursting tenant absorbs the cost.

\evaltakeaway{Per-tenant pacing confines a tenant's burst to its own share: other tenants' misses during a burst drop from 46\% to 15\%.}

\subsection{Co-located Services}
\label{subsection:nv-coloc}
\begin{figure}[t]
  \centering
  \includegraphics[width=\columnwidth]{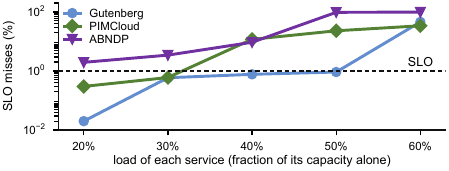}
  \vspace{-0.25in}
  \caption{\textbf{Two co-located services.} U1 and TPC-C (U4) share one server, each at the same fraction of its standalone capacity; SLO misses of the worse service.}
  \label{fig:nv-coloc}
  \vspace{-0.15in}
\end{figure}

Datacenter servers often run several online services at once.
We run the key-value service U1 and TPC-C (U4) on one server.
Each gets half of the CPU workers and half of the cores of each stack, and each runs its own instance of the system under test.
The two share the DRAM, the L2 caches of the stacks, the interconnect and the L3.
We raise both loads together, as a fraction of what \ourname sustains for each service alone on the whole server, and require both services to meet their SLOs (\figref{fig:nv-coloc}).
\ourname serves both up to 50\% of that load, against 30\% for PIMCloud; ABNDP misses the TPC-C SLO already at 20\%.

TPC-C fails first in every system, and for PIMCloud the cause is one stack.
The ten districts of the hottest warehouse all have the same home stack.
Once the CPU workers are busy, PIMCloud sends their requests to the four cores of that stack.
At 40\% load, 87\% of its TPC-C misses occur there, and a missed request waits 977\,$\mu$s for a service of 37\,$\mu$s.
\ourname spreads these districts over the CPU and other stacks as the load rises; a moved district runs about 25\% slower, but its queue disappears.
ABNDP fails for a different reason: its requests themselves slow down.
Sharing the server, it sends 24\% of TPC-C requests away from their home stack, against 14\% when TPC-C runs alone, so they read more data remotely.
Its Stock-Level transactions take 344\,$\mu$s on average, 1.8$\times$ longer than alone.

\evaltakeaway{With two services on one server, \ourname sustains 50\% of each service's standalone load, against 30\% for PIMCloud, by spreading each service's hot data.}

\subsection{Sensitivity}
\label{subsection:nv-sens}

\myparagraph{Skew}
\begin{figure}[t]
  \centering
  \includegraphics[width=\columnwidth]{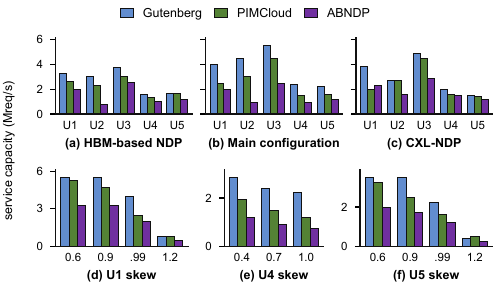}
  \vspace{-0.25in}
  \caption{\textbf{Service capacity across NDP platforms (a--c) and skew (d--f).} (a)~HBM-based NDP; (b)~main configuration; (c)~CXL-attached NDP~\cite{m2ndp}. (d--f)~Zipf exponent of key (U1, U5) or warehouse (U4) popularity.}
  \label{fig:nv-sens}
  \vspace{-0.1in}
\end{figure}
\figref{fig:nv-sens}d--f varies the skew of three services and compares \ourname with PIMCloud and ABNDP.
For key-value services the skew is the Zipf exponent $\alpha$ of key popularity; for TPC-C it is the Zipf exponent of the warehouse that each transaction works on.
\ourname sustains the highest load up to $\alpha = 0.99$ on the key-value services and over the whole range on TPC-C.
The gap widens with the skew: over PIMCloud, it grows from 5\% to 60\% on U1 and from 8\% to 41\% on U5 as $\alpha$ rises from 0.6 to 0.99, and from 1.5$\times$ to 1.9$\times$ on TPC-C.
With more skew, the hot stacks saturate first.
PIMCloud prefers the CPU but falls back to the owner stack once the CPU workers are busy,
whereas \ourname sends cold requests to the CPU before the hot stacks queue up and moves entire hot groups to the CPU or to cold stacks.
Beyond $\alpha = 0.99$, a single group dominates.
At $\alpha = 1.2$, the hottest key-value group carries 19\% of all requests, and its writes run one after another, which caps the capacity of every method at 0.25--0.75\,Mreq/s.

\myparagraph{NDP platform}
\figref{fig:nv-sens}a--c runs the five services on two other NDP platforms.
The HBM-based platform has four NDP cores per stack at 2.4\,GHz, sixteen memory controllers per stack and four CPU ports.
The CXL-NDP platform, modeled after M$^2$NDP~\cite{m2ndp}, has sixteen cores per device at 2\,GHz and two CPU ports of 120\,GB/s, and its devices connect over a 64\,GB/s link.
\ourname sustains the highest load on all five services of the HBM-based platform, 3--33\% above PIMCloud.
On this platform the NDP side is weak, so the CPU must take a larger share of hot loads.
On the CXL-NDP platform, \ourname leads on U1, U3, U4 and U5 by 7--65\% and ties PIMCloud on U2.
Cross-device accesses are costly on this platform, limiting all methods' service capacity on U2, whose requests span devices.

\evaltakeaway{\ourname keeps its lead as skew grows up to $\alpha = 0.99$, and on two other NDP platforms it leads on nine of ten services and ties on the tenth.}

\subsection{Other Services}
\label{subsection:nv-tailbench}

\begin{table}[t]
  \centering
  \caption{TailBench p99 latency (ms) at a fixed load.}
  \label{tab:nv-tailbench}
  \vspace{-0.1in}
  \small
  \setlength{\tabcolsep}{4pt}
  \begin{tabular}{@{}lrrrr@{}}
    \toprule
    Service & \ourname & PIMCloud & ABNDP & CPU-only \\
    \midrule
    Xapian & \textbf{0.86} & 0.88 & 1.26 & 8.49 \\
    Moses & \textbf{12.4} & 13.3 & 29.8 & 12.6 \\
    Sphinx & \textbf{1787} & 2529 & 3380 & 2021 \\
    Img-DNN & \textbf{6.25} & 6.71 & 18.8 & 29.0 \\
    \bottomrule
  \end{tabular}
  \vspace{-0.05in}
\end{table}

\tabref{tab:nv-tailbench} extends the comparison to four TailBench services whose compute and data access differ from the key-value and OLTP services:
search (Xapian), translation (Moses), speech recognition (Sphinx) and image inference (Img-DNN).
\ourname has the lowest p99 on all four services, 2\%, 7\%, 7\% and 29\% below PIMCloud on Xapian, Moses, Img-DNN and Sphinx.
On Sphinx it also beats running every request on the CPU (1.79 vs.\ 2.02\,s), and on Xapian the CPU alone cannot keep up with the load.

\evaltakeaway{On four TailBench services, \ourname has the lowest p99 latency.}

\raggedbottom
\section{Conclusion}
\label{section:conclusion}

We presented \ourname, a management system for online datacenter services on \cpundp servers that combines SLO-aware request routing with coherence- and load-aware data management, and uses drift guards to handle shifting workloads.
Across five services and three NDP platforms, \ourname raises service capacity by up to 65\% (28\% on geometric mean) over the best prior system.

\bibliographystyle{ACM-Reference-Format}
\bibliography{refs}


\begin{thebibliography}{76}

%%% except for \shownote{} and \showURL{}.  The latter does
%%% not use final punctuation, in order to avoid confusing it with
%%% the Web address.

\ifx \showCODEN    \undefined \def \showCODEN     #1{\unskip}     \fi
\ifx \showISBNx    \undefined \def \showISBNx     #1{\unskip}     \fi
\ifx \showISBNxiii \undefined \def \showISBNxiii  #1{\unskip}     \fi
\ifx \showISSN     \undefined \def \showISSN      #1{\unskip}     \fi
\ifx \showLCCN     \undefined \def \showLCCN      #1{\unskip}     \fi
\ifx \shownote     \undefined \def \shownote      #1{#1}          \fi
\ifx \showarticletitle \undefined \def \showarticletitle #1{#1}   \fi
\ifx \showURL      \undefined \def \showURL       {\relax}        \fi
% invisible to TeX
\providecommand\bibfield[2]{#2}
\providecommand\bibinfo[2]{#2}
\providecommand\natexlab[1]{#1}
\providecommand\showeprint[2][]{arXiv:#2}
\makeatletter
\@ifundefined{NAT@parse@date}{}{\let\NAT@parse@date@orig\NAT@parse@date}
\@ifundefined{NAT@parse@date}{}{\def\NAT@parse@date#1#2#3#4#5#6@@{\NAT@parse@date@orig#1#2#3#4#5#6@@\def\NAT@tempyear{0000}\def\NAT@tempexlab{{?}}\ifx\NAT@year\NAT@tempyear\ifx\NAT@exlab\NAT@tempexlab\def\NAT@date{[n.\,d.]}\else\edef\NAT@date{[n.\,d.]\NAT@exlab}\fi\fi}}
\makeatother

\bibitem[Adve et~al\mbox{.}(1991)]%
        {adve1991comparison}
\bibfield{author}{\bibinfo{person}{Sarita~V. Adve}, \bibinfo{person}{Vikram~S.
  Adve}, \bibinfo{person}{Mark~D. Hill}, {and} \bibinfo{person}{Mary~K.
  Vernon}.} \bibinfo{year}{1991}\natexlab{}.
\newblock \showarticletitle{Comparison of Hardware and Software Cache Coherence
  Schemes}. In \bibinfo{booktitle}{\emph{Proceedings of the 18th Annual
  International Symposium on Computer Architecture (ISCA)}}.
  \bibinfo{pages}{298--308}.
\newblock
\href{https://doi.org/10.1145/115952.115982}{doi:\nolinkurl{10.1145/115952.115982}}


\bibitem[Ahn et~al\mbox{.}(2015a)]%
        {ahn2015tesseract}
\bibfield{author}{\bibinfo{person}{Junwhan Ahn}, \bibinfo{person}{Sungpack
  Hong}, \bibinfo{person}{Sungjoo Yoo}, \bibinfo{person}{Onur Mutlu}, {and}
  \bibinfo{person}{Kiyoung Choi}.} \bibinfo{year}{2015}\natexlab{a}.
\newblock \showarticletitle{A scalable processing-in-memory accelerator for
  parallel graph processing}. In \bibinfo{booktitle}{\emph{2015 ACM/IEEE 42nd
  Annual International Symposium on Computer Architecture (ISCA)}}.
  \bibinfo{pages}{105--117}.
\newblock
\href{https://doi.org/10.1145/2749469.2750386}{doi:\nolinkurl{10.1145/2749469.2750386}}


\bibitem[Ahn et~al\mbox{.}(2015b)]%
        {ahn2015piminstructions}
\bibfield{author}{\bibinfo{person}{Junwhan Ahn}, \bibinfo{person}{Sungjoo Yoo},
  \bibinfo{person}{Onur Mutlu}, {and} \bibinfo{person}{Kiyoung Choi}.}
  \bibinfo{year}{2015}\natexlab{b}.
\newblock \showarticletitle{PIM-Enabled Instructions: A Low-Overhead,
  Locality-Aware Processing-in-Memory Architecture}.
\newblock \bibinfo{journal}{\emph{SIGARCH Comput. Archit. News}}
  \bibinfo{volume}{43}, \bibinfo{number}{3S} (\bibinfo{date}{jun}
  \bibinfo{year}{2015}), \bibinfo{pages}{336--348}.
\newblock
\showISSN{0163-5964}
\href{https://doi.org/10.1145/2872887.2750385}{doi:\nolinkurl{10.1145/2872887.2750385}}


\bibitem[Atikoglu et~al\mbox{.}(2012)]%
        {atikoglu2012workload}
\bibfield{author}{\bibinfo{person}{Berk Atikoglu}, \bibinfo{person}{Yuehai Xu},
  \bibinfo{person}{Eitan Frachtenberg}, \bibinfo{person}{Song Jiang}, {and}
  \bibinfo{person}{Mike Paleczny}.} \bibinfo{year}{2012}\natexlab{}.
\newblock \showarticletitle{Workload Analysis of a Large-Scale Key-Value
  Store}. In \bibinfo{booktitle}{\emph{Proceedings of the ACM
  SIGMETRICS/PERFORMANCE Joint International Conference on Measurement and
  Modeling of Computer Systems}}.
\newblock


\bibitem[Barroso et~al\mbox{.}(2018)]%
        {barroso2018datacenter}
\bibfield{author}{\bibinfo{person}{Luiz~Andr{\'e} Barroso},
  \bibinfo{person}{Urs H{\"o}lzle}, {and} \bibinfo{person}{Parthasarathy
  Ranganathan}.} \bibinfo{year}{2018}\natexlab{}.
\newblock \bibinfo{booktitle}{\emph{The Datacenter as a Computer: Designing
  Warehouse-Scale Machines} (\bibinfo{edition}{3} ed.)}.
\newblock \bibinfo{publisher}{Morgan \& Claypool Publishers}.
\newblock
\href{https://doi.org/10.2200/S00874ED3V01Y201809CAC046}{doi:\nolinkurl{10.2200/S00874ED3V01Y201809CAC046}}


\bibitem[Boroumand et~al\mbox{.}(2019)]%
        {boroumand2019conda}
\bibfield{author}{\bibinfo{person}{Amirali Boroumand}, \bibinfo{person}{Saugata
  Ghose}, \bibinfo{person}{Minesh Patel}, \bibinfo{person}{Hasan Hassan},
  \bibinfo{person}{Brandon Lucia}, \bibinfo{person}{Rachata Ausavarungnirun},
  \bibinfo{person}{Kevin Hsieh}, \bibinfo{person}{Nastaran Hajinazar},
  \bibinfo{person}{Krishna~T. Malladi}, \bibinfo{person}{Hongzhong Zheng},
  {and} \bibinfo{person}{Onur Mutlu}.} \bibinfo{year}{2019}\natexlab{}.
\newblock \showarticletitle{CoNDA: Efficient Cache Coherence Support for
  near-Data Accelerators}. In \bibinfo{booktitle}{\emph{Proceedings of the 46th
  International Symposium on Computer Architecture}} (Phoenix, Arizona)
  \emph{(\bibinfo{series}{ISCA '19})}. \bibinfo{publisher}{Association for
  Computing Machinery}, \bibinfo{address}{New York, NY, USA},
  \bibinfo{pages}{629--642}.
\newblock
\showISBNx{9781450366694}
\href{https://doi.org/10.1145/3307650.3322266}{doi:\nolinkurl{10.1145/3307650.3322266}}


\bibitem[Boroumand et~al\mbox{.}(2016)]%
        {boroumand2016lazypim}
\bibfield{author}{\bibinfo{person}{Amirali Boroumand}, \bibinfo{person}{Saugata
  Ghose}, \bibinfo{person}{Minesh Patel}, \bibinfo{person}{Hasan Hassan},
  \bibinfo{person}{Brandon Lucia}, \bibinfo{person}{Kevin Hsieh},
  \bibinfo{person}{Krishna~T Malladi}, \bibinfo{person}{Hongzhong Zheng}, {and}
  \bibinfo{person}{Onur Mutlu}.} \bibinfo{year}{2016}\natexlab{}.
\newblock \showarticletitle{LazyPIM: An efficient cache coherence mechanism for
  processing-in-memory}.
\newblock \bibinfo{journal}{\emph{IEEE Computer Architecture Letters}}
  \bibinfo{volume}{16}, \bibinfo{number}{1} (\bibinfo{year}{2016}),
  \bibinfo{pages}{46--50}.
\newblock


\bibitem[Brin and Page(1998)]%
        {brin1998pagerank}
\bibfield{author}{\bibinfo{person}{Sergey Brin} {and} \bibinfo{person}{Lawrence
  Page}.} \bibinfo{year}{1998}\natexlab{}.
\newblock \showarticletitle{The anatomy of a large-scale hypertextual Web
  search engine}.
\newblock \bibinfo{journal}{\emph{Computer Networks and ISDN Systems}}
  \bibinfo{volume}{30}, \bibinfo{number}{1} (\bibinfo{year}{1998}),
  \bibinfo{pages}{107--117}.
\newblock
\showISSN{0169-7552}
\href{https://doi.org/10.1016/S0169-7552(98)00110-X}{doi:\nolinkurl{10.1016/S0169-7552(98)00110-X}}
\newblock
\shownote{Proceedings of the Seventh International World Wide Web Conference}.


\bibitem[Cao et~al\mbox{.}(2020)]%
        {cao2020rocksdb}
\bibfield{author}{\bibinfo{person}{Zhichao Cao}, \bibinfo{person}{Siying Dong},
  \bibinfo{person}{Sagar Vemuri}, {and} \bibinfo{person}{David H.~C. Du}.}
  \bibinfo{year}{2020}\natexlab{}.
\newblock \showarticletitle{Characterizing, Modeling, and Benchmarking
  {RocksDB} Key-Value Workloads at {Facebook}}. In
  \bibinfo{booktitle}{\emph{Proceedings of the 18th USENIX Conference on File
  and Storage Technologies (FAST)}}. \bibinfo{pages}{209--223}.
\newblock


\bibitem[Chen et~al\mbox{.}(2019)]%
        {chen2019parties}
\bibfield{author}{\bibinfo{person}{Shuang Chen}, \bibinfo{person}{Christina
  Delimitrou}, {and} \bibinfo{person}{Jos\'{e}~F. Mart\'{\i}nez}.}
  \bibinfo{year}{2019}\natexlab{}.
\newblock \showarticletitle{PARTIES: QoS-Aware Resource Partitioning for
  Multiple Interactive Services}. In \bibinfo{booktitle}{\emph{Proceedings of
  the Twenty-Fourth International Conference on Architectural Support for
  Programming Languages and Operating Systems}} (Providence, RI, USA)
  \emph{(\bibinfo{series}{ASPLOS '19})}. \bibinfo{publisher}{Association for
  Computing Machinery}, \bibinfo{address}{New York, NY, USA},
  \bibinfo{pages}{107--120}.
\newblock
\showISBNx{9781450362405}
\href{https://doi.org/10.1145/3297858.3304005}{doi:\nolinkurl{10.1145/3297858.3304005}}


\bibitem[Chen et~al\mbox{.}(2022)]%
        {chen2022pimcloud}
\bibfield{author}{\bibinfo{person}{Shuang Chen}, \bibinfo{person}{Yi Jiang},
  \bibinfo{person}{Christina Delimitrou}, {and} \bibinfo{person}{Jos{\'e}~F.
  Mart{\'\i}nez}.} \bibinfo{year}{2022}\natexlab{}.
\newblock \showarticletitle{{PIMCloud}: {QoS}-Aware Resource Management of
  Latency-Critical Applications in Clouds with Processing-in-Memory}. In
  \bibinfo{booktitle}{\emph{Proceedings of the IEEE International Symposium on
  High-Performance Computer Architecture (HPCA)}}.
\newblock


\bibitem[Chen et~al\mbox{.}(2025)]%
        {chen2025rearchitecting}
\bibfield{author}{\bibinfo{person}{Youmin Chen}, \bibinfo{person}{Jiwu Shu},
  \bibinfo{person}{Yanyan Shen}, \bibinfo{person}{Linpeng Huang}, {and}
  \bibinfo{person}{Hong Mei}.} \bibinfo{year}{2025}\natexlab{}.
\newblock \showarticletitle{Rearchitecting the Thread Model of In-Memory
  Key-Value Stores with uTPS}. In \bibinfo{booktitle}{\emph{Proceedings of the
  ACM SIGOPS 31st Symposium on Operating Systems Principles}} (Lotte Hotel
  World, Seoul, Republic of Korea) \emph{(\bibinfo{series}{SOSP '25})}.
  \bibinfo{publisher}{Association for Computing Machinery},
  \bibinfo{address}{New York, NY, USA}, \bibinfo{pages}{1099--1114}.
\newblock
\showISBNx{9798400718700}
\href{https://doi.org/10.1145/3731569.3764794}{doi:\nolinkurl{10.1145/3731569.3764794}}


\bibitem[Choi et~al\mbox{.}(2025)]%
        {choi2025hera}
\bibfield{author}{\bibinfo{person}{Yujeong Choi}, \bibinfo{person}{John Kim},
  {and} \bibinfo{person}{Minsoo Rhu}.} \bibinfo{year}{2025}\natexlab{}.
\newblock \showarticletitle{Hera: A Heterogeneity-Aware Multi-Tenant Inference
  Server for Personalized Recommendations}. In
  \bibinfo{booktitle}{\emph{Proceedings of the 34th International Conference on
  Parallel Architectures and Compilation Techniques}}
  \emph{(\bibinfo{series}{PACT '25})}. \bibinfo{pages}{320--332}.
\newblock


\bibitem[{Compute Express Link Consortium}(2022)]%
        {cxl3}
\bibfield{author}{\bibinfo{person}{{Compute Express Link Consortium}}.}
  \bibinfo{year}{2022}\natexlab{}.
\newblock \bibinfo{title}{Compute Express Link ({CXL}) Specification, Revision
  3.0}.
\newblock \bibinfo{howpublished}{\url{https://computeexpresslink.org/}}.
\newblock


\bibitem[Cooper et~al\mbox{.}(2010)]%
        {cooper2010benchmarking}
\bibfield{author}{\bibinfo{person}{Brian~F. Cooper}, \bibinfo{person}{Adam
  Silberstein}, \bibinfo{person}{Erwin Tam}, \bibinfo{person}{Raghu
  Ramakrishnan}, {and} \bibinfo{person}{Russell Sears}.}
  \bibinfo{year}{2010}\natexlab{}.
\newblock \showarticletitle{Benchmarking cloud serving systems with YCSB}. In
  \bibinfo{booktitle}{\emph{Proceedings of the 1st ACM Symposium on Cloud
  Computing}} (Indianapolis, Indiana, USA) \emph{(\bibinfo{series}{SoCC '10})}.
  \bibinfo{publisher}{Association for Computing Machinery},
  \bibinfo{address}{New York, NY, USA}, \bibinfo{pages}{143--154}.
\newblock
\showISBNx{9781450300360}
\href{https://doi.org/10.1145/1807128.1807152}{doi:\nolinkurl{10.1145/1807128.1807152}}


\bibitem[Cormode and Muthukrishnan(2005)]%
        {cormode2005countmin}
\bibfield{author}{\bibinfo{person}{Graham Cormode} {and} \bibinfo{person}{S.
  Muthukrishnan}.} \bibinfo{year}{2005}\natexlab{}.
\newblock \showarticletitle{An Improved Data Stream Summary: The Count-Min
  Sketch and its Applications}.
\newblock \bibinfo{journal}{\emph{Journal of Algorithms}} \bibinfo{volume}{55},
  \bibinfo{number}{1} (\bibinfo{year}{2005}), \bibinfo{pages}{58--75}.
\newblock


\bibitem[Curino et~al\mbox{.}(2010)]%
        {curino2010schism}
\bibfield{author}{\bibinfo{person}{Carlo Curino}, \bibinfo{person}{Evan Jones},
  \bibinfo{person}{Yang Zhang}, {and} \bibinfo{person}{Sam Madden}.}
  \bibinfo{year}{2010}\natexlab{}.
\newblock \showarticletitle{Schism: a workload-driven approach to database
  replication and partitioning}.
\newblock \bibinfo{journal}{\emph{Proc. VLDB Endow.}} \bibinfo{volume}{3},
  \bibinfo{number}{1--2} (\bibinfo{date}{Sept.} \bibinfo{year}{2010}),
  \bibinfo{pages}{48--57}.
\newblock
\showISSN{2150-8097}
\href{https://doi.org/10.14778/1920841.1920853}{doi:\nolinkurl{10.14778/1920841.1920853}}


\bibitem[Dean and Barroso(2013)]%
        {dean2013tail}
\bibfield{author}{\bibinfo{person}{Jeffrey Dean} {and}
  \bibinfo{person}{Luiz~Andr\'{e} Barroso}.} \bibinfo{year}{2013}\natexlab{}.
\newblock \showarticletitle{The tail at scale}.
\newblock \bibinfo{journal}{\emph{Commun. ACM}} \bibinfo{volume}{56},
  \bibinfo{number}{2} (\bibinfo{date}{Feb.} \bibinfo{year}{2013}),
  \bibinfo{pages}{74--80}.
\newblock
\showISSN{0001-0782}
\href{https://doi.org/10.1145/2408776.2408794}{doi:\nolinkurl{10.1145/2408776.2408794}}


\bibitem[DeCandia et~al\mbox{.}(2007)]%
        {decandia2007dynamo}
\bibfield{author}{\bibinfo{person}{Giuseppe DeCandia}, \bibinfo{person}{Deniz
  Hastorun}, \bibinfo{person}{Madan Jampani}, \bibinfo{person}{Gunavardhan
  Kakulapati}, \bibinfo{person}{Avinash Lakshman}, \bibinfo{person}{Alex
  Pilchin}, \bibinfo{person}{Swaminathan Sivasubramanian},
  \bibinfo{person}{Peter Vosshall}, {and} \bibinfo{person}{Werner Vogels}.}
  \bibinfo{year}{2007}\natexlab{}.
\newblock \showarticletitle{Dynamo: amazon's highly available key-value store}.
  In \bibinfo{booktitle}{\emph{Proceedings of Twenty-First ACM SIGOPS Symposium
  on Operating Systems Principles}} (Stevenson, Washington, USA)
  \emph{(\bibinfo{series}{SOSP '07})}. \bibinfo{publisher}{Association for
  Computing Machinery}, \bibinfo{address}{New York, NY, USA},
  \bibinfo{pages}{205--220}.
\newblock
\showISBNx{9781595935915}
\href{https://doi.org/10.1145/1294261.1294281}{doi:\nolinkurl{10.1145/1294261.1294281}}


\bibitem[{DragonflyDB}(2026)]%
        {dragonflydb}
\bibfield{author}{\bibinfo{person}{{DragonflyDB}}.}
  \bibinfo{year}{2026}\natexlab{}.
\newblock \bibinfo{title}{Dragonfly: A Modern Replacement for {Redis} and
  {Memcached}}.
\newblock
  \bibinfo{howpublished}{\url{https://github.com/dragonflydb/dragonfly}}.
\newblock
\newblock
\shownote{Accessed: 2026-07-27}.


\bibitem[Einziger et~al\mbox{.}(2017)]%
        {einziger2017tinylfu}
\bibfield{author}{\bibinfo{person}{Gil Einziger}, \bibinfo{person}{Roy
  Friedman}, {and} \bibinfo{person}{Ben Manes}.}
  \bibinfo{year}{2017}\natexlab{}.
\newblock \showarticletitle{{TinyLFU}: A Highly Efficient Cache Admission
  Policy}.
\newblock \bibinfo{journal}{\emph{ACM Transactions on Storage}}
  \bibinfo{volume}{13}, \bibinfo{number}{4} (\bibinfo{year}{2017}).
\newblock


\bibitem[Fried et~al\mbox{.}(2020)]%
        {fried2020caladan}
\bibfield{author}{\bibinfo{person}{Joshua Fried}, \bibinfo{person}{Zhenyuan
  Ruan}, \bibinfo{person}{Amy Ousterhout}, {and} \bibinfo{person}{Adam Belay}.}
  \bibinfo{year}{2020}\natexlab{}.
\newblock \showarticletitle{Caladan: Mitigating Interference at Microsecond
  Timescales}. In \bibinfo{booktitle}{\emph{Proceedings of the 14th USENIX
  Symposium on Operating Systems Design and Implementation (OSDI)}}.
  \bibinfo{pages}{281--297}.
\newblock


\bibitem[Goder et~al\mbox{.}(2015)]%
        {goder2015bistro}
\bibfield{author}{\bibinfo{person}{Andrey Goder}, \bibinfo{person}{Alexey
  Spiridonov}, {and} \bibinfo{person}{Yin Wang}.}
  \bibinfo{year}{2015}\natexlab{}.
\newblock \showarticletitle{Bistro: Scheduling {Data-Parallel} Jobs Against
  Live Production Systems}. In \bibinfo{booktitle}{\emph{2015 USENIX Annual
  Technical Conference (USENIX ATC 15)}}. \bibinfo{publisher}{USENIX
  Association}, \bibinfo{address}{Santa Clara, CA}, \bibinfo{pages}{459--471}.
\newblock
\showISBNx{978-1-931971-225}
\urldef\tempurl%
\url{https://www.usenix.org/conference/atc15/technical-session/presentation/goder}
\showURL{%
\tempurl}


\bibitem[Gupta et~al\mbox{.}(2020)]%
        {gupta2020recommendation}
\bibfield{author}{\bibinfo{person}{Udit Gupta}, \bibinfo{person}{Carole-Jean
  Wu}, \bibinfo{person}{Xiaodong Wang}, \bibinfo{person}{Maxim Naumov},
  \bibinfo{person}{Brandon Reagen}, \bibinfo{person}{David Brooks},
  \bibinfo{person}{Bradford Cottel}, \bibinfo{person}{Kim Hazelwood},
  \bibinfo{person}{Mark Hempstead}, \bibinfo{person}{Bill Jia},
  \bibinfo{person}{Hsien-Hsin~S. Lee}, \bibinfo{person}{Andrey Malevich},
  \bibinfo{person}{Dheevatsa Mudigere}, \bibinfo{person}{Mikhail Smelyanskiy},
  \bibinfo{person}{Liang Xiong}, {and} \bibinfo{person}{Xuan Zhang}.}
  \bibinfo{year}{2020}\natexlab{}.
\newblock \showarticletitle{The Architectural Implications of Facebook's
  DNN-Based Personalized Recommendation}. In
  \bibinfo{booktitle}{\emph{Proceedings of the IEEE International Symposium on
  High Performance Computer Architecture}} \emph{(\bibinfo{series}{HPCA '20})}.
  \bibinfo{pages}{488--501}.
\newblock
\href{https://doi.org/10.1109/HPCA47549.2020.00047}{doi:\nolinkurl{10.1109/HPCA47549.2020.00047}}


\bibitem[Gómez-Luna et~al\mbox{.}(2022)]%
        {gomez2022benchmarking}
\bibfield{author}{\bibinfo{person}{Juan Gómez-Luna}, \bibinfo{person}{Izzat~El
  Hajj}, \bibinfo{person}{Ivan Fernandez}, \bibinfo{person}{Christina
  Giannoula}, \bibinfo{person}{Geraldo~F. Oliveira}, {and}
  \bibinfo{person}{Onur Mutlu}.} \bibinfo{year}{2022}\natexlab{}.
\newblock \showarticletitle{Benchmarking a New Paradigm: Experimental Analysis
  and Characterization of a Real Processing-in-Memory System}.
\newblock \bibinfo{journal}{\emph{IEEE Access}}  \bibinfo{volume}{10}
  (\bibinfo{year}{2022}), \bibinfo{pages}{52565--52608}.
\newblock
\href{https://doi.org/10.1109/ACCESS.2022.3174101}{doi:\nolinkurl{10.1109/ACCESS.2022.3174101}}


\bibitem[Ham et~al\mbox{.}(2024)]%
        {m2ndp}
\bibfield{author}{\bibinfo{person}{Hyungkyu Ham}, \bibinfo{person}{Jeongmin
  Hong}, \bibinfo{person}{Geonwoo Park}, \bibinfo{person}{Yunseon Shin},
  \bibinfo{person}{Okkyun Woo}, \bibinfo{person}{Wonhyuk Yang},
  \bibinfo{person}{Jinhoon Bae}, \bibinfo{person}{Eunhyeok Park},
  \bibinfo{person}{Hyojin Sung}, \bibinfo{person}{Euicheol Lim}, {and}
  \bibinfo{person}{Gwangsun Kim}.} \bibinfo{year}{2024}\natexlab{}.
\newblock \showarticletitle{Low-overhead General-purpose Near-Data Processing
  in {CXL} Memory Expanders}. In \bibinfo{booktitle}{\emph{Proceedings of the
  57th IEEE/ACM International Symposium on Microarchitecture (MICRO)}}.
\newblock


\bibitem[Haque et~al\mbox{.}(2015)]%
        {haque2015few}
\bibfield{author}{\bibinfo{person}{Md~E. Haque}, \bibinfo{person}{Yong~hun
  Eom}, \bibinfo{person}{Yuxiong He}, \bibinfo{person}{Sameh Elnikety},
  \bibinfo{person}{Ricardo Bianchini}, {and} \bibinfo{person}{Kathryn~S.
  McKinley}.} \bibinfo{year}{2015}\natexlab{}.
\newblock \showarticletitle{Few-to-Many: Incremental Parallelism for Reducing
  Tail Latency in Interactive Services}.
\newblock \bibinfo{journal}{\emph{SIGARCH Comput. Archit. News}}
  \bibinfo{volume}{43}, \bibinfo{number}{1} (\bibinfo{date}{March}
  \bibinfo{year}{2015}), \bibinfo{pages}{161--175}.
\newblock
\showISSN{0163-5964}
\href{https://doi.org/10.1145/2786763.2694384}{doi:\nolinkurl{10.1145/2786763.2694384}}


\bibitem[Hsieh et~al\mbox{.}(2016)]%
        {hsieh2016tom}
\bibfield{author}{\bibinfo{person}{Kevin Hsieh}, \bibinfo{person}{Eiman
  Ebrahimi}, \bibinfo{person}{Gwangsun Kim}, \bibinfo{person}{Niladrish
  Chatterjee}, \bibinfo{person}{Mike O'Connor}, \bibinfo{person}{Nandita
  Vijaykumar}, \bibinfo{person}{Onur Mutlu}, {and} \bibinfo{person}{Stephen~W.
  Keckler}.} \bibinfo{year}{2016}\natexlab{}.
\newblock \showarticletitle{Transparent Offloading and Mapping ({TOM}):
  Enabling Programmer-Transparent Near-Data Processing in {GPU} Systems}. In
  \bibinfo{booktitle}{\emph{Proceedings of the 43rd International Symposium on
  Computer Architecture (ISCA)}}. \bibinfo{pages}{204--216}.
\newblock


\bibitem[Hsu et~al\mbox{.}(2015)]%
        {hsu2015adrenaline}
\bibfield{author}{\bibinfo{person}{Chang-Hong Hsu}, \bibinfo{person}{Yunqi
  Zhang}, \bibinfo{person}{Michael~A. Laurenzano}, \bibinfo{person}{David
  Meisner}, \bibinfo{person}{Thomas Wenisch}, \bibinfo{person}{Jason Mars},
  \bibinfo{person}{Lingjia Tang}, {and} \bibinfo{person}{Ronald~G.
  Dreslinski}.} \bibinfo{year}{2015}\natexlab{}.
\newblock \showarticletitle{Adrenaline: Pinpointing and reining in tail queries
  with quick voltage boosting}. In \bibinfo{booktitle}{\emph{2015 IEEE 21st
  International Symposium on High Performance Computer Architecture (HPCA)}}.
  \bibinfo{pages}{271--282}.
\newblock
\href{https://doi.org/10.1109/HPCA.2015.7056039}{doi:\nolinkurl{10.1109/HPCA.2015.7056039}}


\bibitem[Huang et~al\mbox{.}(2025)]%
        {huang2025tigon}
\bibfield{author}{\bibinfo{person}{Yibo Huang}, \bibinfo{person}{Haowei Chen},
  \bibinfo{person}{Newton Ni}, \bibinfo{person}{Yan Sun},
  \bibinfo{person}{Vijay Chidambaram}, \bibinfo{person}{Dixin Tang}, {and}
  \bibinfo{person}{Emmett Witchel}.} \bibinfo{year}{2025}\natexlab{}.
\newblock \showarticletitle{Tigon: A Distributed Database for a {CXL} Pod}. In
  \bibinfo{booktitle}{\emph{19th USENIX Symposium on Operating Systems Design
  and Implementation (OSDI 25)}}. \bibinfo{publisher}{USENIX Association},
  \bibinfo{address}{Boston, MA}, \bibinfo{pages}{109--128}.
\newblock
\showISBNx{978-1-939133-47-2}
\urldef\tempurl%
\url{https://www.usenix.org/conference/osdi25/presentation/huang-yibo}
\showURL{%
\tempurl}


\bibitem[Huggins-Daines et~al\mbox{.}(2006)]%
        {huggins2006pocketsphinx}
\bibfield{author}{\bibinfo{person}{David Huggins-Daines},
  \bibinfo{person}{Mohit Kumar}, \bibinfo{person}{Arthur Chan},
  \bibinfo{person}{Alan~W. Black}, \bibinfo{person}{Mosur Ravishankar}, {and}
  \bibinfo{person}{Alexander~I. Rudnicky}.} \bibinfo{year}{2006}\natexlab{}.
\newblock \showarticletitle{{PocketSphinx}: A Free, Real-Time Continuous Speech
  Recognition System for Hand-Held Devices}. In
  \bibinfo{booktitle}{\emph{Proceedings of the IEEE International Conference on
  Acoustics, Speech and Signal Processing (ICASSP)}}.
\newblock


\bibitem[Jeddeloh and Keeth(2012)]%
        {jeddeloh2012hybriddram}
\bibfield{author}{\bibinfo{person}{Joe Jeddeloh} {and} \bibinfo{person}{Brent
  Keeth}.} \bibinfo{year}{2012}\natexlab{}.
\newblock \showarticletitle{Hybrid memory cube new DRAM architecture increases
  density and performance}. In \bibinfo{booktitle}{\emph{2012 Symposium on VLSI
  Technology (VLSIT)}}. \bibinfo{pages}{87--88}.
\newblock
\href{https://doi.org/10.1109/VLSIT.2012.6242474}{doi:\nolinkurl{10.1109/VLSIT.2012.6242474}}


\bibitem[Jin et~al\mbox{.}(2017)]%
        {jin2017netcache}
\bibfield{author}{\bibinfo{person}{Xin Jin}, \bibinfo{person}{Xiaozhou Li},
  \bibinfo{person}{Haoyu Zhang}, \bibinfo{person}{Robert Soul{\'e}},
  \bibinfo{person}{Jeongkeun Lee}, \bibinfo{person}{Nate Foster},
  \bibinfo{person}{Changhoon Kim}, {and} \bibinfo{person}{Ion Stoica}.}
  \bibinfo{year}{2017}\natexlab{}.
\newblock \showarticletitle{NetCache: Balancing Key-Value Stores with Fast
  In-Network Caching}. In \bibinfo{booktitle}{\emph{Proceedings of the 26th
  Symposium on Operating Systems Principles (SOSP '17)}}.
  \bibinfo{pages}{121--136}.
\newblock
\href{https://doi.org/10.1145/3132747.3132764}{doi:\nolinkurl{10.1145/3132747.3132764}}


\bibitem[Jouppi et~al\mbox{.}(2017)]%
        {jouppi2017datacenter}
\bibfield{author}{\bibinfo{person}{Norman~P. Jouppi}, \bibinfo{person}{Cliff
  Young}, \bibinfo{person}{Nishant Patil}, \bibinfo{person}{David Patterson},
  \bibinfo{person}{Gaurav Agrawal}, \bibinfo{person}{Raminder Bajwa},
  \bibinfo{person}{Sarah Bates}, \bibinfo{person}{Suresh Bhatia},
  \bibinfo{person}{Nan Boden}, \bibinfo{person}{Al Borchers},
  \bibinfo{person}{Rick Boyle}, \bibinfo{person}{Pierre-luc Cantin},
  \bibinfo{person}{Clifford Chao}, \bibinfo{person}{Chris Clark},
  \bibinfo{person}{Jeremy Coriell}, \bibinfo{person}{Mike Daley},
  \bibinfo{person}{Matt Dau}, \bibinfo{person}{Jeffrey Dean},
  \bibinfo{person}{Ben Gelb}, \bibinfo{person}{Tara~Vazir Ghaemmaghami},
  \bibinfo{person}{Rajendra Gottipati}, \bibinfo{person}{William Gulland},
  \bibinfo{person}{Robert Hagmann}, \bibinfo{person}{C.~Richard Ho},
  \bibinfo{person}{Doug Hogberg}, \bibinfo{person}{John Hu},
  \bibinfo{person}{Robert Hundt}, \bibinfo{person}{Dan Hurt},
  \bibinfo{person}{Julian Ibarz}, \bibinfo{person}{Aaron Jaffey},
  \bibinfo{person}{Alek Jaworski}, \bibinfo{person}{Alexander Kaplan},
  \bibinfo{person}{Harshit Khaitan}, \bibinfo{person}{Daniel Killebrew},
  \bibinfo{person}{Andy Koch}, \bibinfo{person}{Naveen Kumar},
  \bibinfo{person}{Steve Lacy}, \bibinfo{person}{James Laudon},
  \bibinfo{person}{James Law}, \bibinfo{person}{Diemthu Le},
  \bibinfo{person}{Chris Leary}, \bibinfo{person}{Zhuyuan Liu},
  \bibinfo{person}{Kyle Lucke}, \bibinfo{person}{Alan Lundin},
  \bibinfo{person}{Gordon MacKean}, \bibinfo{person}{Adriana Maggiore},
  \bibinfo{person}{Maire Mahony}, \bibinfo{person}{Kieran Miller},
  \bibinfo{person}{Rahul Nagarajan}, \bibinfo{person}{Ravi Narayanaswami},
  \bibinfo{person}{Ray Ni}, \bibinfo{person}{Kathy Nix},
  \bibinfo{person}{Thomas Norrie}, \bibinfo{person}{Mark Omernick},
  \bibinfo{person}{Narayana Penukonda}, \bibinfo{person}{Andy Phelps},
  \bibinfo{person}{Jonathan Ross}, \bibinfo{person}{Matt Ross},
  \bibinfo{person}{Amir Salek}, \bibinfo{person}{Emad Samadiani},
  \bibinfo{person}{Chris Severn}, \bibinfo{person}{Gregory Sizikov},
  \bibinfo{person}{Matthew Snelham}, \bibinfo{person}{Jed Souter},
  \bibinfo{person}{Dan Steinberg}, \bibinfo{person}{Andy Swing},
  \bibinfo{person}{Mercedes Tan}, \bibinfo{person}{Gregory Thorson},
  \bibinfo{person}{Bo Tian}, \bibinfo{person}{Horia Toma},
  \bibinfo{person}{Erick Tuttle}, \bibinfo{person}{Vijay Vasudevan},
  \bibinfo{person}{Richard Walter}, \bibinfo{person}{Walter Wang},
  \bibinfo{person}{Eric Wilcox}, {and} \bibinfo{person}{Doe~Hyun Yoon}.}
  \bibinfo{year}{2017}\natexlab{}.
\newblock \showarticletitle{In-Datacenter Performance Analysis of a Tensor
  Processing Unit}.
\newblock \bibinfo{journal}{\emph{SIGARCH Comput. Archit. News}}
  \bibinfo{volume}{45}, \bibinfo{number}{2} (\bibinfo{date}{June}
  \bibinfo{year}{2017}), \bibinfo{pages}{1--12}.
\newblock
\showISSN{0163-5964}
\href{https://doi.org/10.1145/3140659.3080246}{doi:\nolinkurl{10.1145/3140659.3080246}}


\bibitem[Kaffes et~al\mbox{.}(2019)]%
        {kaffes2019shinjuku}
\bibfield{author}{\bibinfo{person}{Kostis Kaffes}, \bibinfo{person}{Timothy
  Chong}, \bibinfo{person}{Jack~Tigar Humphries}, \bibinfo{person}{Adam Belay},
  \bibinfo{person}{David Mazi{\`e}res}, {and} \bibinfo{person}{Christos
  Kozyrakis}.} \bibinfo{year}{2019}\natexlab{}.
\newblock \showarticletitle{Shinjuku: Preemptive Scheduling for
  {${\mu}$second-scale} Tail Latency}. In \bibinfo{booktitle}{\emph{16th USENIX
  Symposium on Networked Systems Design and Implementation (NSDI 19)}}.
  \bibinfo{publisher}{USENIX Association}, \bibinfo{address}{Boston, MA},
  \bibinfo{pages}{345--360}.
\newblock
\showISBNx{978-1-931971-49-2}
\urldef\tempurl%
\url{https://www.usenix.org/conference/nsdi19/presentation/kaffes}
\showURL{%
\tempurl}


\bibitem[Kang et~al\mbox{.}(2022)]%
        {kang2022pimtreevldb}
\bibfield{author}{\bibinfo{person}{Hongbo Kang}, \bibinfo{person}{Yiwei Zhao},
  \bibinfo{person}{Guy~E. Blelloch}, \bibinfo{person}{Laxman Dhulipala},
  \bibinfo{person}{Yan Gu}, \bibinfo{person}{Charles McGuffey}, {and}
  \bibinfo{person}{Phillip~B. Gibbons}.} \bibinfo{year}{2022}\natexlab{}.
\newblock \showarticletitle{PIM-Tree: A Skew-Resistant Index for
  Processing-in-Memory}.
\newblock \bibinfo{journal}{\emph{Proc. VLDB Endow.}} \bibinfo{volume}{16},
  \bibinfo{number}{4} (\bibinfo{date}{dec} \bibinfo{year}{2022}),
  \bibinfo{pages}{946--958}.
\newblock
\showISSN{2150-8097}
\href{https://doi.org/10.14778/3574245.3574275}{doi:\nolinkurl{10.14778/3574245.3574275}}


\bibitem[Kasture and Sanchez(2016)]%
        {kasture2016tailbench}
\bibfield{author}{\bibinfo{person}{Harshad Kasture} {and}
  \bibinfo{person}{Daniel Sanchez}.} \bibinfo{year}{2016}\natexlab{}.
\newblock \showarticletitle{{TailBench}: A Benchmark Suite and Evaluation
  Methodology for Latency-Critical Applications}. In
  \bibinfo{booktitle}{\emph{Proceedings of the IEEE International Symposium on
  Workload Characterization (IISWC)}}.
\newblock


\bibitem[Kemper and Neumann(2011)]%
        {kemper2011hyper}
\bibfield{author}{\bibinfo{person}{Alfons Kemper} {and} \bibinfo{person}{Thomas
  Neumann}.} \bibinfo{year}{2011}\natexlab{}.
\newblock \showarticletitle{HyPer: A hybrid OLTP\&OLAP main memory database
  system based on virtual memory snapshots}. In \bibinfo{booktitle}{\emph{2011
  IEEE 27th International Conference on Data Engineering}}.
  \bibinfo{pages}{195--206}.
\newblock
\href{https://doi.org/10.1109/ICDE.2011.5767867}{doi:\nolinkurl{10.1109/ICDE.2011.5767867}}


\bibitem[Kim et~al\mbox{.}(2016)]%
        {kim2016ramulator}
\bibfield{author}{\bibinfo{person}{Yoongu Kim}, \bibinfo{person}{Weikun Yang},
  {and} \bibinfo{person}{Onur Mutlu}.} \bibinfo{year}{2016}\natexlab{}.
\newblock \showarticletitle{Ramulator: A Fast and Extensible {DRAM} Simulator}.
\newblock \bibinfo{journal}{\emph{IEEE Computer Architecture Letters}}
  \bibinfo{volume}{15}, \bibinfo{number}{1} (\bibinfo{year}{2016}),
  \bibinfo{pages}{45--49}.
\newblock


\bibitem[Koehn et~al\mbox{.}(2007)]%
        {koehn2007moses}
\bibfield{author}{\bibinfo{person}{Philipp Koehn}, \bibinfo{person}{Hieu
  Hoang}, \bibinfo{person}{Alexandra Birch}, \bibinfo{person}{Chris
  Callison-Burch}, \bibinfo{person}{Marcello Federico}, \bibinfo{person}{Nicola
  Bertoldi}, \bibinfo{person}{Brooke Cowan}, \bibinfo{person}{Wade Shen},
  \bibinfo{person}{Christine Moran}, \bibinfo{person}{Richard Zens},
  \bibinfo{person}{Chris Dyer}, \bibinfo{person}{Ond{\v{r}}ej Bojar},
  \bibinfo{person}{Alexandra Constantin}, {and} \bibinfo{person}{Evan Herbst}.}
  \bibinfo{year}{2007}\natexlab{}.
\newblock \showarticletitle{Moses: Open Source Toolkit for Statistical Machine
  Translation}. In \bibinfo{booktitle}{\emph{Proceedings of the 45th Annual
  Meeting of the Association for Computational Linguistics (ACL), Demo and
  Poster Sessions}}. \bibinfo{pages}{177--180}.
\newblock


\bibitem[Kwon et~al\mbox{.}(2021)]%
        {kwon2021samsunghbmpim}
\bibfield{author}{\bibinfo{person}{Young-Cheon Kwon}, \bibinfo{person}{Suk~Han
  Lee}, \bibinfo{person}{Jaehoon Lee}, \bibinfo{person}{Sang-Hyuk Kwon},
  \bibinfo{person}{Je~Min Ryu}, \bibinfo{person}{Jong-Pil Son},
  \bibinfo{person}{O Seongil}, \bibinfo{person}{Hak-Soo Yu},
  \bibinfo{person}{Haesuk Lee}, \bibinfo{person}{Soo~Young Kim},
  \bibinfo{person}{Youngmin Cho}, \bibinfo{person}{Jin~Guk Kim},
  \bibinfo{person}{Jongyoon Choi}, \bibinfo{person}{Hyun-Sung Shin},
  \bibinfo{person}{Jin Kim}, \bibinfo{person}{BengSeng Phuah},
  \bibinfo{person}{HyoungMin Kim}, \bibinfo{person}{Myeong~Jun Song},
  \bibinfo{person}{Ahn Choi}, \bibinfo{person}{Daeho Kim},
  \bibinfo{person}{SooYoung Kim}, \bibinfo{person}{Eun-Bong Kim},
  \bibinfo{person}{David Wang}, \bibinfo{person}{Shinhaeng Kang},
  \bibinfo{person}{Yuhwan Ro}, \bibinfo{person}{Seungwoo Seo},
  \bibinfo{person}{JoonHo Song}, \bibinfo{person}{Jaeyoun Youn},
  \bibinfo{person}{Kyomin Sohn}, {and} \bibinfo{person}{Nam~Sung Kim}.}
  \bibinfo{year}{2021}\natexlab{}.
\newblock \showarticletitle{25.4 A 20nm 6GB Function-In-Memory DRAM, Based on
  HBM2 with a 1.2TFLOPS Programmable Computing Unit Using Bank-Level
  Parallelism, for Machine Learning Applications}. In
  \bibinfo{booktitle}{\emph{2021 IEEE International Solid- State Circuits
  Conference (ISSCC)}}, Vol.~\bibinfo{volume}{64}. \bibinfo{pages}{350--352}.
\newblock


\bibitem[Lee et~al\mbox{.}(2014)]%
        {lee2014hbm}
\bibfield{author}{\bibinfo{person}{Dong~Uk Lee}, \bibinfo{person}{Kyung~Whan
  Kim}, \bibinfo{person}{Kwan~Weon Kim}, \bibinfo{person}{Hongjung Kim},
  \bibinfo{person}{Ju~Young Kim}, \bibinfo{person}{Young~Jun Park},
  \bibinfo{person}{Jae~Hwan Kim}, \bibinfo{person}{Dae~Suk Kim},
  \bibinfo{person}{Heat~Bit Park}, \bibinfo{person}{Jin~Wook Shin},
  \bibinfo{person}{Jang~Hwan Cho}, \bibinfo{person}{Ki~Hun Kwon},
  \bibinfo{person}{Min~Jeong Kim}, \bibinfo{person}{Jaejin Lee},
  \bibinfo{person}{Kun~Woo Park}, \bibinfo{person}{Byongtae Chung}, {and}
  \bibinfo{person}{Sungjoo Hong}.} \bibinfo{year}{2014}\natexlab{}.
\newblock \showarticletitle{25.2 A 1.2V 8Gb 8-channel 128GB/s high-bandwidth
  memory (HBM) stacked DRAM with effective microbump I/O test methods using
  29nm process and TSV}. In \bibinfo{booktitle}{\emph{2014 IEEE International
  Solid-State Circuits Conference Digest of Technical Papers (ISSCC)}}.
  \bibinfo{pages}{432--433}.
\newblock
\href{https://doi.org/10.1109/ISSCC.2014.6757501}{doi:\nolinkurl{10.1109/ISSCC.2014.6757501}}


\bibitem[Li et~al\mbox{.}(2020)]%
        {li2020pegasus}
\bibfield{author}{\bibinfo{person}{Jialin Li}, \bibinfo{person}{Jacob Nelson},
  \bibinfo{person}{Ellis Michael}, \bibinfo{person}{Xin Jin}, {and}
  \bibinfo{person}{Dan R.~K. Ports}.} \bibinfo{year}{2020}\natexlab{}.
\newblock \showarticletitle{Pegasus: Tolerating Skewed Workloads in Distributed
  Storage with {In-Network} Coherence Directories}. In
  \bibinfo{booktitle}{\emph{14th USENIX Symposium on Operating Systems Design
  and Implementation (OSDI 20)}}. \bibinfo{publisher}{USENIX Association},
  \bibinfo{pages}{387--406}.
\newblock
\showISBNx{978-1-939133-19-9}
\urldef\tempurl%
\url{https://www.usenix.org/conference/osdi20/presentation/li-jialin}
\showURL{%
\tempurl}


\bibitem[Lim et~al\mbox{.}(2017)]%
        {lim2017cicada}
\bibfield{author}{\bibinfo{person}{Hyeontaek Lim}, \bibinfo{person}{Michael
  Kaminsky}, {and} \bibinfo{person}{David~G. Andersen}.}
  \bibinfo{year}{2017}\natexlab{}.
\newblock \showarticletitle{Cicada: Dependably Fast Multi-Core In-Memory
  Transactions}. In \bibinfo{booktitle}{\emph{Proceedings of the 2017 ACM
  International Conference on Management of Data}} (Chicago, Illinois, USA)
  \emph{(\bibinfo{series}{SIGMOD '17})}. \bibinfo{publisher}{Association for
  Computing Machinery}, \bibinfo{address}{New York, NY, USA},
  \bibinfo{pages}{21--35}.
\newblock
\showISBNx{9781450341974}
\href{https://doi.org/10.1145/3035918.3064015}{doi:\nolinkurl{10.1145/3035918.3064015}}


\bibitem[Lin et~al\mbox{.}(2026)]%
        {lin2026atumai}
\bibfield{author}{\bibinfo{person}{Qiushi Lin}, \bibinfo{person}{Chaojie
  Zhang}, \bibinfo{person}{Íñigo Goiri}, \bibinfo{person}{Aditya Akella},
  \bibinfo{person}{Ricardo Bianchini}, {and} \bibinfo{person}{Jovan
  Stojkovic}.} \bibinfo{year}{2026}\natexlab{}.
\newblock \bibinfo{title}{AtumAI: A Principled Framework for Agentic Generation
  of Datacenter Control-Plane Policies}.
\newblock
\showeprint[arxiv]{2608.02569}~[cs.AI]
\urldef\tempurl%
\url{https://arxiv.org/abs/2608.02569}
\showURL{%
\tempurl}


\bibitem[Lo et~al\mbox{.}(2015)]%
        {lo2015heracles}
\bibfield{author}{\bibinfo{person}{David Lo}, \bibinfo{person}{Liqun Cheng},
  \bibinfo{person}{Rama Govindaraju}, \bibinfo{person}{Parthasarathy
  Ranganathan}, {and} \bibinfo{person}{Christos Kozyrakis}.}
  \bibinfo{year}{2015}\natexlab{}.
\newblock \showarticletitle{Heracles: Improving Resource Efficiency at Scale}.
  In \bibinfo{booktitle}{\emph{Proceedings of the 42nd Annual International
  Symposium on Computer Architecture}} \emph{(\bibinfo{series}{ISCA '15})}.
\newblock
\href{https://doi.org/10.1145/2749469.2749475}{doi:\nolinkurl{10.1145/2749469.2749475}}


\bibitem[Lozi et~al\mbox{.}(2012)]%
        {lozi2012rcl}
\bibfield{author}{\bibinfo{person}{Jean-Pierre Lozi}, \bibinfo{person}{Florian
  David}, \bibinfo{person}{Ga{\"e}l Thomas}, \bibinfo{person}{Julia Lawall},
  {and} \bibinfo{person}{Gilles Muller}.} \bibinfo{year}{2012}\natexlab{}.
\newblock \showarticletitle{Remote Core Locking: Migrating Critical-Section
  Execution to Improve the Performance of Multithreaded Applications}. In
  \bibinfo{booktitle}{\emph{Proceedings of the USENIX Annual Technical
  Conference (ATC)}}. \bibinfo{pages}{65--76}.
\newblock


\bibitem[Mao et~al\mbox{.}(2012)]%
        {mao2012masstree}
\bibfield{author}{\bibinfo{person}{Yandong Mao}, \bibinfo{person}{Eddie
  Kohler}, {and} \bibinfo{person}{Robert~Tappan Morris}.}
  \bibinfo{year}{2012}\natexlab{}.
\newblock \showarticletitle{Cache Craftiness for Fast Multicore Key-Value
  Storage}. In \bibinfo{booktitle}{\emph{Proceedings of the 7th ACM European
  Conference on Computer Systems (EuroSys)}}.
\newblock


\bibitem[Metwally et~al\mbox{.}(2005)]%
        {metwally2005spacesaving}
\bibfield{author}{\bibinfo{person}{Ahmed Metwally}, \bibinfo{person}{Divyakant
  Agrawal}, {and} \bibinfo{person}{Amr~El Abbadi}.}
  \bibinfo{year}{2005}\natexlab{}.
\newblock \showarticletitle{Efficient Computation of Frequent and Top-k
  Elements in Data Streams}. In \bibinfo{booktitle}{\emph{Proceedings of the
  10th International Conference on Database Theory (ICDT)}}.
\newblock


\bibitem[Mitzenmacher(2001)]%
        {mitzenmacher2001power}
\bibfield{author}{\bibinfo{person}{Michael Mitzenmacher}.}
  \bibinfo{year}{2001}\natexlab{}.
\newblock \showarticletitle{The Power of Two Choices in Randomized Load
  Balancing}.
\newblock \bibinfo{journal}{\emph{IEEE Transactions on Parallel and Distributed
  Systems}} \bibinfo{volume}{12}, \bibinfo{number}{10} (\bibinfo{year}{2001}),
  \bibinfo{pages}{1094--1104}.
\newblock


\bibitem[Mohoney et~al\mbox{.}(2025)]%
        {mohoney2025quake}
\bibfield{author}{\bibinfo{person}{Jason Mohoney}, \bibinfo{person}{Devesh
  Sarda}, \bibinfo{person}{Mengze Tang}, \bibinfo{person}{Shihabur~Rahman
  Chowdhury}, \bibinfo{person}{Anil Pacaci}, \bibinfo{person}{Ihab~F. Ilyas},
  \bibinfo{person}{Theodoros Rekatsinas}, {and} \bibinfo{person}{Shivaram
  Venkataraman}.} \bibinfo{year}{2025}\natexlab{}.
\newblock \showarticletitle{Quake: Adaptive Indexing for Vector Search}. In
  \bibinfo{booktitle}{\emph{19th USENIX Symposium on Operating Systems Design
  and Implementation (OSDI 25)}}. \bibinfo{publisher}{USENIX Association},
  \bibinfo{address}{Boston, MA}, \bibinfo{pages}{153--169}.
\newblock
\showISBNx{978-1-939133-47-2}
\urldef\tempurl%
\url{https://www.usenix.org/conference/osdi25/presentation/mohoney}
\showURL{%
\tempurl}


\bibitem[Moina-Rivera et~al\mbox{.}(2024)]%
        {moina2024cloudvideo}
\bibfield{author}{\bibinfo{person}{Wilmer Moina-Rivera},
  \bibinfo{person}{Miguel Garcia-Pineda}, \bibinfo{person}{Juan
  Guti{\'e}rrez-Aguado}, {and} \bibinfo{person}{Jose~M. Alcaraz-Calero}.}
  \bibinfo{year}{2024}\natexlab{}.
\newblock \showarticletitle{Cloud Media Video Encoding: Review and Challenges}.
\newblock \bibinfo{journal}{\emph{Multimedia Tools and Applications}}
  \bibinfo{volume}{83} (\bibinfo{year}{2024}), \bibinfo{pages}{81231--81278}.
\newblock
\href{https://doi.org/10.1007/s11042-024-18763-2}{doi:\nolinkurl{10.1007/s11042-024-18763-2}}


\bibitem[Mutlu et~al\mbox{.}(2023)]%
        {mutlu2023primer}
\bibfield{author}{\bibinfo{person}{Onur Mutlu}, \bibinfo{person}{Saugata
  Ghose}, \bibinfo{person}{Juan G{\'o}mez-Luna}, {and} \bibinfo{person}{Rachata
  Ausavarungnirun}.} \bibinfo{year}{2023}\natexlab{}.
\newblock \bibinfo{booktitle}{\emph{A Modern Primer on Processing in Memory}}.
\newblock \bibinfo{publisher}{Springer Nature Singapore},
  \bibinfo{address}{Singapore}, \bibinfo{pages}{171--243}.
\newblock
\showISBNx{978-981-16-7487-7}
\href{https://doi.org/10.1007/978-981-16-7487-7_7}{doi:\nolinkurl{10.1007/978-981-16-7487-7_7}}


\bibitem[Nishtala et~al\mbox{.}(2013)]%
        {nishtala2013scaling}
\bibfield{author}{\bibinfo{person}{Rajesh Nishtala}, \bibinfo{person}{Hans
  Fugal}, \bibinfo{person}{Steven Grimm}, \bibinfo{person}{Marc Kwiatkowski},
  \bibinfo{person}{Herman Lee}, \bibinfo{person}{Harry~C. Li},
  \bibinfo{person}{Ryan McElroy}, \bibinfo{person}{Mike Paleczny},
  \bibinfo{person}{Daniel Peek}, \bibinfo{person}{Paul Saab},
  \bibinfo{person}{David Stafford}, \bibinfo{person}{Tony Tung}, {and}
  \bibinfo{person}{Venkateshwaran Venkataramani}.}
  \bibinfo{year}{2013}\natexlab{}.
\newblock \showarticletitle{Scaling Memcache at Facebook}. In
  \bibinfo{booktitle}{\emph{Proceedings of the 10th USENIX Symposium on
  Networked Systems Design and Implementation (NSDI)}}.
  \bibinfo{pages}{385--398}.
\newblock
\urldef\tempurl%
\url{https://www.usenix.org/conference/nsdi13/technical-sessions/presentation/nishtala}
\showURL{%
\tempurl}


\bibitem[Ousterhout et~al\mbox{.}(2019)]%
        {ousterhout2019shenango}
\bibfield{author}{\bibinfo{person}{Amy Ousterhout}, \bibinfo{person}{Joshua
  Fried}, \bibinfo{person}{Jonathan Behrens}, \bibinfo{person}{Adam Belay},
  {and} \bibinfo{person}{Hari Balakrishnan}.} \bibinfo{year}{2019}\natexlab{}.
\newblock \showarticletitle{Shenango: achieving high CPU efficiency for
  latency-sensitive datacenter workloads}. In
  \bibinfo{booktitle}{\emph{Proceedings of the 16th USENIX Conference on
  Networked Systems Design and Implementation}} (Boston, MA, USA)
  \emph{(\bibinfo{series}{NSDI'19})}. \bibinfo{publisher}{USENIX Association},
  \bibinfo{address}{USA}, \bibinfo{pages}{361–377}.
\newblock
\showISBNx{9781931971492}


\bibitem[Qi et~al\mbox{.}(2025)]%
        {qi2025finbench}
\bibfield{author}{\bibinfo{person}{Shipeng Qi}, \bibinfo{person}{Lingwei Guo},
  \bibinfo{person}{Yuan Lu}, \bibinfo{person}{Chao Zhang},
  \bibinfo{person}{Dong Yang}, \bibinfo{person}{Guanyu Chen},
  \bibinfo{person}{Xinyi Liu}, \bibinfo{person}{Chuan Shen},
  \bibinfo{person}{Yubo Xu}, \bibinfo{person}{Guang Li}, {et~al\mbox{.}}}
  \bibinfo{year}{2025}\natexlab{}.
\newblock \showarticletitle{The {LDBC} Financial Benchmark: Transaction
  Workload}.
\newblock \bibinfo{journal}{\emph{Proceedings of the VLDB Endowment}}
  \bibinfo{volume}{18}, \bibinfo{number}{9} (\bibinfo{year}{2025}),
  \bibinfo{pages}{3007--3020}.
\newblock
\href{https://doi.org/10.14778/3746405.3746424}{doi:\nolinkurl{10.14778/3746405.3746424}}


\bibitem[Qiu et~al\mbox{.}(2018)]%
        {qiu2018real}
\bibfield{author}{\bibinfo{person}{Xiafei Qiu}, \bibinfo{person}{Wubin Cen},
  \bibinfo{person}{Zhengping Qian}, \bibinfo{person}{You Peng},
  \bibinfo{person}{Ying Zhang}, \bibinfo{person}{Xuemin Lin}, {and}
  \bibinfo{person}{Jingren Zhou}.} \bibinfo{year}{2018}\natexlab{}.
\newblock \showarticletitle{Real-time constrained cycle detection in large
  dynamic graphs}.
\newblock \bibinfo{journal}{\emph{Proc. VLDB Endow.}} \bibinfo{volume}{11},
  \bibinfo{number}{12} (\bibinfo{date}{Aug.} \bibinfo{year}{2018}),
  \bibinfo{pages}{1876--1888}.
\newblock
\showISSN{2150-8097}
\href{https://doi.org/10.14778/3229863.3229874}{doi:\nolinkurl{10.14778/3229863.3229874}}


\bibitem[{Redis}(2026)]%
        {redis}
\bibfield{author}{\bibinfo{person}{{Redis}}.} \bibinfo{year}{2026}\natexlab{}.
\newblock \bibinfo{title}{Redis}.
\newblock \bibinfo{howpublished}{\url{https://redis.io/}}.
\newblock
\newblock
\shownote{Accessed November 2026}.


\bibitem[Roghanchi et~al\mbox{.}(2017)]%
        {roghanchi2017ffwd}
\bibfield{author}{\bibinfo{person}{Sepideh Roghanchi}, \bibinfo{person}{Jakob
  Eriksson}, {and} \bibinfo{person}{Nilanjana Basu}.}
  \bibinfo{year}{2017}\natexlab{}.
\newblock \showarticletitle{ffwd: Delegation is (Much) Faster Than You Think}.
  In \bibinfo{booktitle}{\emph{Proceedings of the 26th ACM Symposium on
  Operating Systems Principles (SOSP)}}.
\newblock


\bibitem[Sanchez and Kozyrakis(2013)]%
        {sanchez2013zsim}
\bibfield{author}{\bibinfo{person}{Daniel Sanchez} {and}
  \bibinfo{person}{Christos Kozyrakis}.} \bibinfo{year}{2013}\natexlab{}.
\newblock \showarticletitle{{ZSim}: Fast and Accurate Microarchitectural
  Simulation of Thousand-Core Systems}. In
  \bibinfo{booktitle}{\emph{Proceedings of the 40th Annual International
  Symposium on Computer Architecture (ISCA)}}.
\newblock


\bibitem[{SK hynix}(2022)]%
        {skhynix2022gddr6aim}
\bibfield{author}{\bibinfo{person}{{SK hynix}}.}
  \bibinfo{year}{2022}\natexlab{}.
\newblock \bibinfo{title}{{SK hynix} Develops {PIM}, Next-Generation {AI}
  Accelerator}.
\newblock
  \bibinfo{howpublished}{\url{https://news.skhynix.com/sk-hynix-develops-pim-next-generation-ai-accelerator/}}.
\newblock
\newblock
\shownote{Accessed: 2026-07-10}.


\bibitem[Tarjan(1975)]%
        {tarjan1975union}
\bibfield{author}{\bibinfo{person}{Robert~Endre Tarjan}.}
  \bibinfo{year}{1975}\natexlab{}.
\newblock \showarticletitle{Efficiency of a Good But Not Linear Set Union
  Algorithm}.
\newblock \bibinfo{journal}{\emph{J. ACM}} \bibinfo{volume}{22},
  \bibinfo{number}{2} (\bibinfo{year}{1975}), \bibinfo{pages}{215--225}.
\newblock


\bibitem[Teguia et~al\mbox{.}(2024)]%
        {teguia2024vpim}
\bibfield{author}{\bibinfo{person}{Dufy Teguia}, \bibinfo{person}{Jiaxuan
  Chen}, \bibinfo{person}{Stella Bitchebe}, \bibinfo{person}{Oana Balmau},
  {and} \bibinfo{person}{Alain Tchana}.} \bibinfo{year}{2024}\natexlab{}.
\newblock \showarticletitle{vPIM: Processing-in-Memory Virtualization}. In
  \bibinfo{booktitle}{\emph{Proceedings of the 25th International Middleware
  Conference}} (Hong Kong, Hong Kong) \emph{(\bibinfo{series}{Middleware
  '24})}. \bibinfo{publisher}{Association for Computing Machinery},
  \bibinfo{address}{New York, NY, USA}, \bibinfo{pages}{417–430}.
\newblock
\showISBNx{9798400706233}
\href{https://doi.org/10.1145/3652892.3700782}{doi:\nolinkurl{10.1145/3652892.3700782}}


\bibitem[Tian et~al\mbox{.}(2023)]%
        {tian2023abndp}
\bibfield{author}{\bibinfo{person}{Boyu Tian}, \bibinfo{person}{Qihang Chen},
  {and} \bibinfo{person}{Mingyu Gao}.} \bibinfo{year}{2023}\natexlab{}.
\newblock \showarticletitle{{ABNDP}: Co-optimizing Data Access and Load Balance
  in Near-Data Processing}. In \bibinfo{booktitle}{\emph{Proceedings of the
  28th ACM International Conference on Architectural Support for Programming
  Languages and Operating Systems (ASPLOS)}}.
\newblock


\bibitem[{Transaction Processing Performance Council}(2010)]%
        {tpcc}
\bibfield{author}{\bibinfo{person}{{Transaction Processing Performance
  Council}}.} \bibinfo{year}{2010}\natexlab{}.
\newblock \bibinfo{title}{{TPC} Benchmark {C} Standard Specification, Revision
  5.11}.
\newblock \bibinfo{howpublished}{\url{https://www.tpc.org/tpcc/}}.
\newblock


\bibitem[Tu et~al\mbox{.}(2013)]%
        {tu2013silo}
\bibfield{author}{\bibinfo{person}{Stephen Tu}, \bibinfo{person}{Wenting
  Zheng}, \bibinfo{person}{Eddie Kohler}, \bibinfo{person}{Barbara Liskov},
  {and} \bibinfo{person}{Samuel Madden}.} \bibinfo{year}{2013}\natexlab{}.
\newblock \showarticletitle{Speedy Transactions in Multicore In-Memory
  Databases}. In \bibinfo{booktitle}{\emph{Proceedings of the 24th ACM
  Symposium on Operating Systems Principles (SOSP)}}.
\newblock


\bibitem[UPMEM(2026)]%
        {upmem}
\bibfield{author}{\bibinfo{person}{UPMEM}.} \bibinfo{year}{2026}\natexlab{}.
\newblock \bibinfo{title}{UPMEM Technology}.
\newblock \bibinfo{howpublished}{\url{https://www.upmem.com/technology/}}.
\newblock
\newblock
\shownote{Accessed July 2026}.


\bibitem[Xapian(2026)]%
        {xapian}
\bibfield{author}{\bibinfo{person}{Xapian}.} \bibinfo{year}{2026}\natexlab{}.
\newblock \bibinfo{title}{Xapian project}.
\newblock \bibinfo{howpublished}{\url{https://xapian.org/}}.
\newblock
\newblock
\shownote{Accessed November 2026}.


\bibitem[{XCENA}(2026)]%
        {xcena2026mx1}
\bibfield{author}{\bibinfo{person}{{XCENA}}.} \bibinfo{year}{2026}\natexlab{}.
\newblock \bibinfo{title}{{MX1}: {CXL} Computational Memory}.
\newblock \bibinfo{howpublished}{\url{https://xcena.com/about}}.
\newblock
\newblock
\shownote{Accessed: 2026-07-10}.


\bibitem[Xia et~al\mbox{.}(2025)]%
        {xia2025modm}
\bibfield{author}{\bibinfo{person}{Yuchen Xia}, \bibinfo{person}{Divyam
  Sharma}, \bibinfo{person}{Yichao Yuan}, \bibinfo{person}{Souvik Kundu}, {and}
  \bibinfo{person}{Nishil Talati}.} \bibinfo{year}{2025}\natexlab{}.
\newblock \showarticletitle{MoDM: Efficient Serving for Image Generation via
  Mixture-of-Diffusion Models}. In \bibinfo{booktitle}{\emph{Proceedings of the
  31st ACM International Conference on Architectural Support for Programming
  Languages and Operating Systems, Volume 1}} (Pittsburgh, PA, USA)
  \emph{(\bibinfo{series}{ASPLOS '26})}. \bibinfo{publisher}{Association for
  Computing Machinery}, \bibinfo{address}{New York, NY, USA},
  \bibinfo{pages}{163--182}.
\newblock
\showISBNx{9798400721656}
\href{https://doi.org/10.1145/3760250.3762220}{doi:\nolinkurl{10.1145/3760250.3762220}}


\bibitem[Yang et~al\mbox{.}(2020)]%
        {yang2020twitter}
\bibfield{author}{\bibinfo{person}{Juncheng Yang}, \bibinfo{person}{Yao Yue},
  {and} \bibinfo{person}{K.~V. Rashmi}.} \bibinfo{year}{2020}\natexlab{}.
\newblock \showarticletitle{A Large Scale Analysis of Hundreds of In-memory
  Cache Clusters at {Twitter}}. In \bibinfo{booktitle}{\emph{Proceedings of the
  14th USENIX Symposium on Operating Systems Design and Implementation
  (OSDI)}}.
\newblock


\bibitem[Yang et~al\mbox{.}(2021)]%
        {yang2021large}
\bibfield{author}{\bibinfo{person}{Juncheng Yang}, \bibinfo{person}{Yao Yue},
  {and} \bibinfo{person}{K.~V. Rashmi}.} \bibinfo{year}{2021}\natexlab{}.
\newblock \showarticletitle{A Large-scale Analysis of Hundreds of In-memory
  Key-value Cache Clusters at Twitter}.
\newblock \bibinfo{journal}{\emph{ACM Trans. Storage}} \bibinfo{volume}{17},
  \bibinfo{number}{3}, Article \bibinfo{articleno}{17} (\bibinfo{date}{Aug.}
  \bibinfo{year}{2021}), \bibinfo{numpages}{35}~pages.
\newblock
\showISSN{1553-3077}
\href{https://doi.org/10.1145/3468521}{doi:\nolinkurl{10.1145/3468521}}


\bibitem[Yuan(2015)]%
        {yuan_multilayer_convnet}
\bibfield{author}{\bibinfo{person}{Xingdi Yuan}.}
  \bibinfo{year}{2015}\natexlab{}.
\newblock \bibinfo{title}{{multi-layer-convnet}}.
\newblock \bibinfo{howpublished}{GitHub repository}.
\newblock
\urldef\tempurl%
\url{https://github.com/xingdi-eric-yuan/multi-layer-convnet}
\showURL{%
\tempurl}


\bibitem[Zhang et~al\mbox{.}(2023)]%
        {zhang2023shepherd}
\bibfield{author}{\bibinfo{person}{Hong Zhang}, \bibinfo{person}{Yupeng Tang},
  \bibinfo{person}{Anurag Khandelwal}, {and} \bibinfo{person}{Ion Stoica}.}
  \bibinfo{year}{2023}\natexlab{}.
\newblock \showarticletitle{{SHEPHERD}: Serving {DNNs} in the Wild}. In
  \bibinfo{booktitle}{\emph{20th USENIX Symposium on Networked Systems Design
  and Implementation (NSDI 23)}}. \bibinfo{publisher}{USENIX Association},
  \bibinfo{address}{Boston, MA}, \bibinfo{pages}{787--808}.
\newblock
\showISBNx{978-1-939133-33-5}
\urldef\tempurl%
\url{https://www.usenix.org/conference/nsdi23/presentation/zhang-hong}
\showURL{%
\tempurl}


\bibitem[Zhang et~al\mbox{.}(2018)]%
        {zhang2018graphp}
\bibfield{author}{\bibinfo{person}{Mingxing Zhang}, \bibinfo{person}{Youwei
  Zhuo}, \bibinfo{person}{Chao Wang}, \bibinfo{person}{Mingyu Gao},
  \bibinfo{person}{Yongwei Wu}, \bibinfo{person}{Kang Chen},
  \bibinfo{person}{Christos Kozyrakis}, {and} \bibinfo{person}{Xuehai Qian}.}
  \bibinfo{year}{2018}\natexlab{}.
\newblock \showarticletitle{GraphP: Reducing Communication for PIM-Based Graph
  Processing with Efficient Data Partition}. In \bibinfo{booktitle}{\emph{2018
  IEEE International Symposium on High Performance Computer Architecture
  (HPCA)}}. \bibinfo{pages}{544--557}.
\newblock
\href{https://doi.org/10.1109/HPCA.2018.00053}{doi:\nolinkurl{10.1109/HPCA.2018.00053}}


\bibitem[Zipf(1949)]%
        {zipf2016human}
\bibfield{author}{\bibinfo{person}{George~Kingsley Zipf}.}
  \bibinfo{year}{1949}\natexlab{}.
\newblock \bibinfo{booktitle}{\emph{Human Behavior and the Principle of Least
  Effort: An Introduction to Human Ecology}}.
\newblock \bibinfo{publisher}{Addison-Wesley Press},
  \bibinfo{address}{Cambridge, MA}.
\newblock
\urldef\tempurl%
\url{https://www.mpi.nl/publications/item2407822/human-behavior-and-principle-least-effort-introduction-human-eoclogy}
\showURL{%
\tempurl}


\end{thebibliography}

\end{document}